\documentclass[twocolumn,amsmath,amssymb,amssymb,aps,pra, showpacs]{revtex4-1}
\usepackage{graphicx,epsfig,epstopdf}% Include figure files
\usepackage{mathrsfs}
\usepackage{dcolumn}% Align table columns on decimal point
\usepackage{bm}% bold math
\usepackage{braket}
\usepackage{color}
\begin{document}
\title{Collective resonances in $\chi^{(3)}$; a QED study}
\author{Konstantin E. Dorfman}
\email{Email: kdorfman@uci.edu}
\author{Shaul Mukamel}
\affiliation{University of California, Irvine, California 92697-2025}
\date{\today}
\pacs{}    

\begin{abstract} 

We calculate  the third order susceptibility $\chi^{(3)}$ of a pair of 2 level atoms which interact via the exchange of photons. QED corrections to second order in coupling to vacuum field modes yield collective two photon absorption resonances which can be observed in transmission spectroscopy of shaped broadband pulses. While some collective effects  can be obtained by introducing an effective interatomic coupling  using a quantum master equation, the predicted signals contain clear features that are missed by that level of theory and require a full diagrammatic QED  treatment.
\end{abstract} 
                                   % Activate to display a given date or no date
\maketitle

\section{Introduction}

Many-body effects strongly influence electronic and optical properties of atoms, molecules and materials \cite{Bas11, Kul11, Ore12,Hau04,Mah11}. Collective resonances which involve several particles in multidimensional spectroscopy  \cite{Muk95} provide clear signatures of these effects. Delocalized excitons play a key role in the function of light harvesting antennae and reaction centers \cite{Che09,Col10,Dor131,Gro00}. Quantum information processing schemes \cite{Nie00} have been proposed that exploit collective resonances due to long-range dipole-dipole coupling \cite{Luk01}. 
%One technique is two-dimensional double-quantum coherence optical signals \cite{Sto09,Dai12} which involve several interactions with classical and quantum modes of the radiation field\cite{Ric11,Coh97}. 

In this paper we use a diagrammatic approach to calculate the transmission of a broadband pulse to fourth order in coupling to classical modes and second order in the coupling to quantum vacuum modes. Calculations are made  in the joint field plus matter space  \cite{Salam} starting with the multipolar hamiltonian \cite{Pow80, Pow82,Cra84,Salam} where all couplings are mediated by the exchange of photons. We find QED  contributions to the semiclassical (SC) susceptibility $\chi^{(3)}$ that originate from mixed time ordering of interactions with vacuum and classical modes. These result in collective resonances that can be probed by shaped broadband pulses.

Shaped femtosecond pulses \cite{Wol07,Wal09} has been used extensively as a tool for coherent control of the fundamental processes in various systems \cite{Pes07,Wan10,Roy10}. The combination of narrow and broadband laser fields have been used in excited state CARS measurements \cite{Xin02,Mul02,Kee04,Vol05,Vac07,Wal09}. We find that such pulses can suppress the background of the single-particle resonances and highlight the collective resonances. The phase profile of the broadband field relative to narrowband provides a coherent control tool for manipulating the collective resonances.
 
  \begin{figure}[t]
\begin{center}
\includegraphics[trim=0cm 0cm 0cm 0cm,angle=0, width=0.4\textwidth]{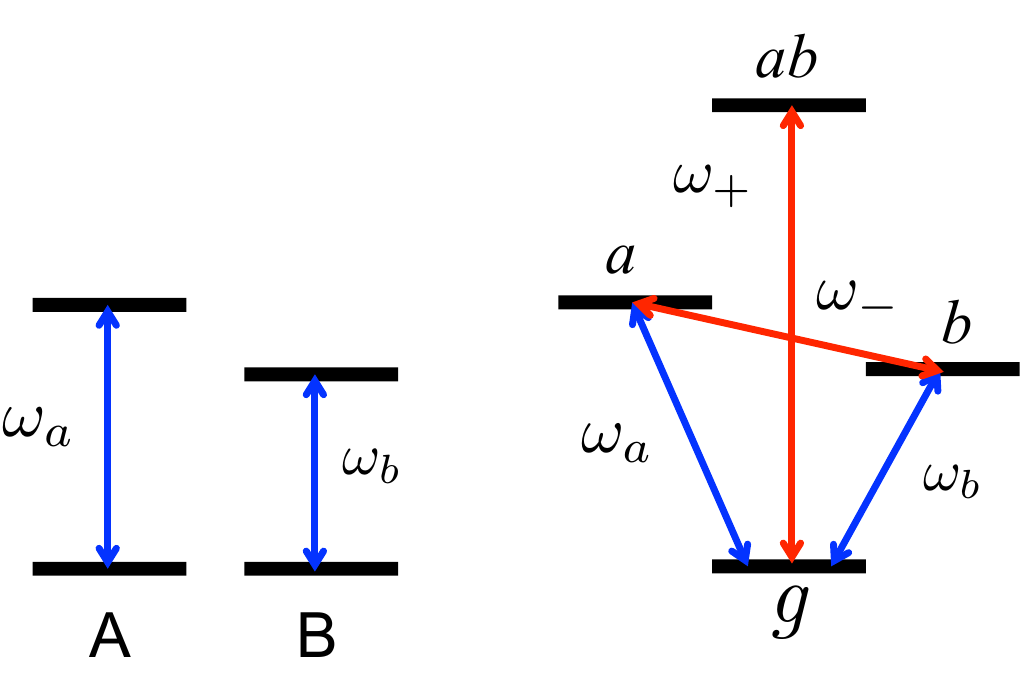}
\end{center}
\caption{(Color online) The two noninteracting atoms A and B - left and corresponding two-particle eigenstates - right. Blue arrows represent single particle resonances with individual atom, red arrows correspond to collective two-particle resonance with frequency $\omega_+$ - sum frequency and $\omega_-$ - difference frequency.}
\label{fig:scheme}
\end{figure}

The response of a quantum system to classical optical fields is  commonly described by semiclassical (SC) susceptibilities \cite{Muk95}. These are calculated by sums over states of matter. Spontaneous emission is included either phenomenologically or via a quantum master equation (QME) \cite{Aga74,Scu97}. The predicted new collective resonances are induced by weak coupling to vacuum modes \cite{Ric11} which causes QED corrections to SC susceptibilities. The QME provides an approximate description of QED effects and can only partially account for these resonances.

\section{The Hamiltonian}

The multipolar Hamiltonian for two systems $A$ and $B$ and the radiation field is given by \cite{Cra84,Salam}$H=H_0+H_{int}$, $H_0$ is the unperturbed Hamiltonian 
\begin{equation}
H_0=H_A+H_B+H_F,
\end{equation}
where the matter Hamiltonian reads
\begin{align}
H_A+H_B=\hbar\omega_A\hat{\sigma}_{A}^{(z)}+\hbar\omega_B\hat{\sigma}_{B}^{(z)},
\end{align}
and $\sigma^{(z)}$ are Pauli matrices. The field Hamiltonian is 
\begin{align}
H_F=\frac{1}{2}\int d\mathbf{r}[\epsilon_0|\hat{\mathbf{E}}(\mathbf{r})|^2+\mu_0|\hat{\mathbf{H}}(\mathbf{r})|^2].
\end{align}
The field-matter interaction in the rotating wave approximation written in the interaction picture with respect to $H_0$ is
\begin{align}\label{eq:Hint}
H_{int}(t)=\int d\mathbf{r}\hat{\mathbf{E}}^{\dagger}(t,\mathbf{r})\hat{\mathbf{V}}(t,\mathbf{r})+H.c,
\end{align}
where $\hat{\mathbf{V}}(t,\mathbf{r})=\sum_{\alpha}\hat{\mathbf{V}}^{\alpha}(t)\delta(\mathbf{r}-\mathbf{r}_{\alpha})$ is a matter operator representing the lowering (exciton annihilation) part of the dipole coupling  and $\alpha$ run over atoms located at $\mathbf{r}_{\alpha}$. The field operator is 
\begin{equation}\label{eq:Ejs}
\hat{\mathbf{E}}(t,\mathbf{r})=\sum_{\mathbf{k}_s,\mu}\left(\frac{2\pi\hbar \omega_s}{\Omega}\right)^{1/2}\epsilon^{(\mu)}(\mathbf{k}_s)\hat{a}_{\mathbf{k}_s}e^{-i\omega_st+i\mathbf{k}_s\cdot\mathbf{r}},
\end{equation}
where $\epsilon^{(\mu)}(\mathbf{k})$ is the unit electric polarization vector of mode $(\mathbf{k}_s,\mu)$, $\mu$ being the index of polarization, $\omega_s=c|\mathbf{k}_s|$, $c$ is speed of light, $\Omega$ is quantization volume. For classical field modes (represented by e.g. coherent state) we can replace the field operator by its expectation value  $\mathbf{E}(t,\mathbf{r})=\langle \psi|\hat{\mathbf{E}}(t,\mathbf{r})|\psi\rangle$, where $\psi$ represent the state of light. Otherwise we treat it as an operator. We shall make use of the following commutation relations \cite{Salam}
\begin{align}\label{eq:com}
[E^{(l)}(\tau_i,\mathbf{r}_\beta),E^{(m)\dagger}(\tau_j,\mathbf{r}_\alpha)]=\int\frac{d\omega}{2\pi}\mathcal{D}_{\alpha\beta}^{(l,m)}(\omega)e^{i\omega(\tau_j-\tau_i)}, 
\end{align}
\begin{align}\label{eq:com1}
[E^{(l)}(\omega_i,\mathbf{r}_\beta),E^{(m)\dagger}(\omega_j,\mathbf{r}_\alpha)]=\mathcal{D}_{\alpha\beta}^{(l,m)}(\omega_i)\delta(\omega_i-\omega_j), 
\end{align}
where $l$ and $m$ denote cartesian components of the electric field, and the coupling tensor reads \cite{Salam, Dor13} 
\begin{align}\label{eq:Dab}
\mathcal{D}_{\alpha\beta}^{(l,m)}(\omega)=\frac{\hbar}{2\pi\epsilon_0}(-\mathbf{\nabla}^2\delta_{lm}+\nabla_l\cdot\nabla_m)\frac{\sin\left(\omega r_{\alpha\beta}/c\right)}{r_{\alpha\beta}},
\end{align}
and $r_{\alpha\beta}=|\mathbf{r}_\alpha-\mathbf{r}_\beta|$ is the interatomic distance. The diagonal elements $\mathcal{D}_{\alpha\alpha}^{(l,m)}(\omega)=\hbar\omega^3/2\pi\epsilon_0c^3\delta_{lm}$ represent the self energy corrections - energy shifts and cooperative emission rate whereas the off diagonal contribution (\ref{eq:Dab}) yields the cross relaxation and dipole-dipole coupling.

When classical light interacts with ensemble of noninteracting atoms $A$ and $B$, the response is additive and is given by 
$S(\omega)=S_A(\omega)+S_B(\omega)$ where $S_A(S_B)$ are individual responses of each atom. For weakly-coupled atoms the response acquires non-additive terms $S_{AB}(\omega)$ which arise from interactions between atoms
\begin{equation}\label{eq:nadd}
S(\omega)=S_A(\omega)+S_B(\omega)+S_{AB}(\omega).
\end{equation}

We shall calculate these non additive contributions perturbatively in light/matter interactions using QED and show that they contain two types of collective resonances: two-photon absorption (TPA) $\omega+\omega_1=\omega_A+\omega_B$ and Raman type $\omega-\omega_1=\omega_A-\omega_B$  where $\omega$ and $\omega_1$ are two field modes of the transmitted pulse and $\omega_A$, $\omega_B$ are transition frequencies of the two atoms $A$ and $B$, respectively. We show that the signal may not be fully described by an effective Hamiltonian alone but a complete QED treatment is needed. We further show how such resonances may be observed and distinguished from non collective single-particle resonances.

  \section{Transmission of a broadband pulse to second order in coupling to vacuum modes}

  \begin{figure*}[t]
\begin{center}
\includegraphics[trim=0cm 0cm 0cm 0cm,angle=0, width=0.65\textwidth]{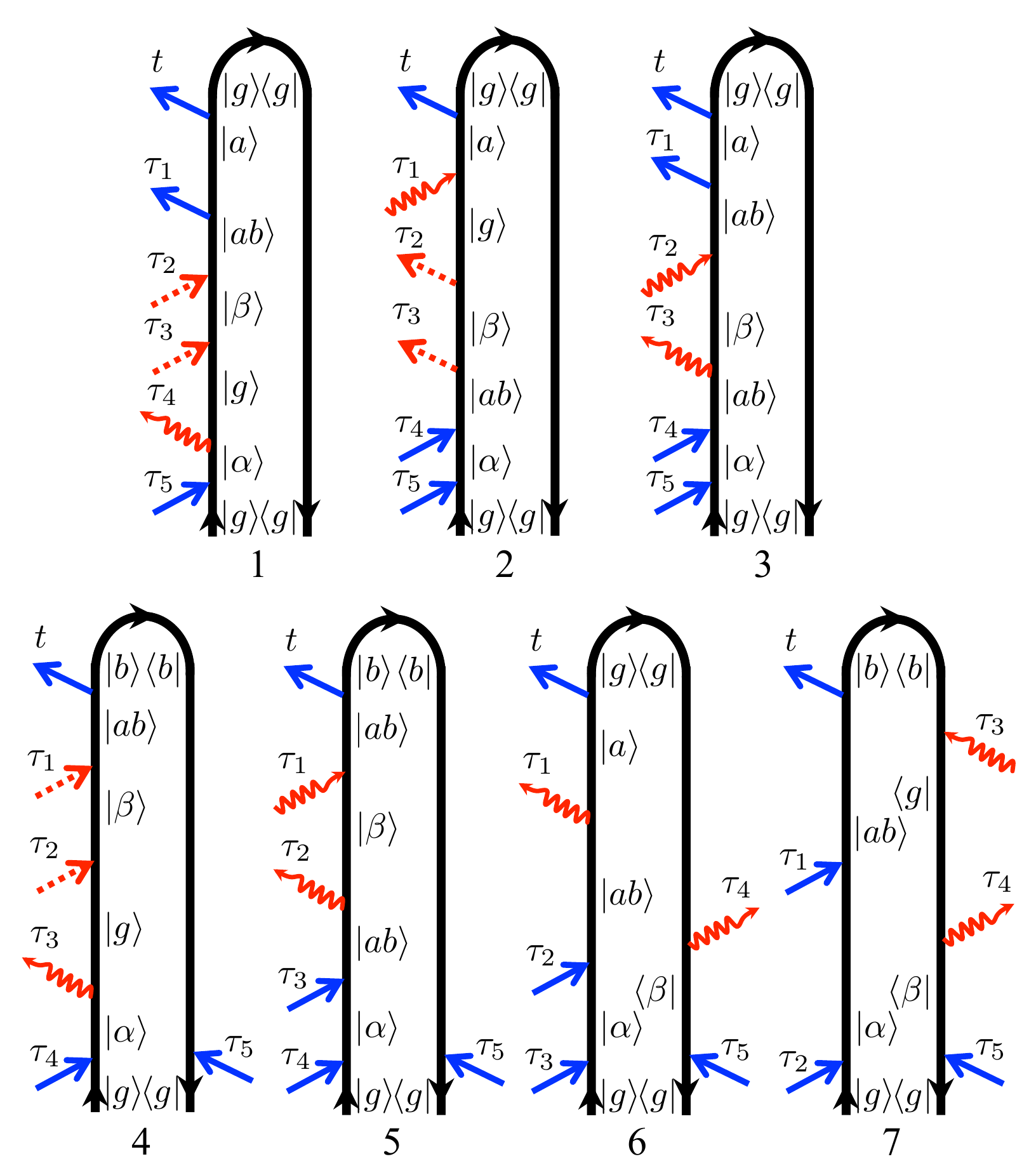}
\end{center}
\caption{(Color online) Loop diagrams (for rules see \cite{Rah10}) for the frequency dispersed transmission signal (\ref{eq:S0}) from a pair of noninteracting atoms $A$ and $B$, $\alpha,\beta=a,b$ that are initially prepared in the ground state $g$ generated by four interactions with classical light and two - with quantum field modes.  Shown are diagrams where the last interaction is with atom $A$. Interchanging $A$ and $B$ will yield another set of diagrams. Straight blue arrows represent field-matter interaction of classical light  with atoms. Wavy red lines correspond to quantum modes, dashed red lines represent interactions with both classical and quantum modes.}
\label{fig:6g}
\end{figure*}

We assume that the system interacts with a classical broadband shaped pulse $E(\omega,\mathbf{r})=\int_{-\infty}^{\infty}dtE(t)e^{i\omega t-i\mathbf{k}_0\cdot\mathbf{r}}$, where we assume that all frequency components of the incoming pulse have the same wave vector $\mathbf{k}_0$ (paraxial approximation). We focus on the frequency-dispersed transmission
\begin{equation}\label{eq:S0}
S(\omega)=\frac{2}{\hbar}\int_{-\infty}^{\infty}d\mathbf{r}\text{Im}[E^{*}(\omega,\mathbf{r})P(\omega,\mathbf{r})],
\end{equation}
where 
\begin{equation}
P(\omega,\mathbf{r})=\int_{-\infty}^{\infty}dtP(t,\mathbf{r})e^{i\omega t}
\end{equation}
is the Fourier transform of the polarization. We shall calculate $P$ perturbatively in the field-matter interaction (Eq. (\ref{eq:Hint})).  To maintain a convenient bookkeeping of time ordered Green's functions we adopt superoperator notation. With every ordinary operator $A$ we associate two superoperators defined by their action on an ordinary operator $X$ as $A_L=AX$  acting from left, $A_R=XA$ (right). We further define the symmetric and antisymmetric combinations $A_+=\frac{1}{2}(A_L+A_R)$, $A_-=A_L-A_R$. Without loss of generality we assume that the last interaction results in deexcitation of the matter with consequent emission of the photon and express the nonlinear polarization using superoperators in the interaction picture
\begin{equation}\label{eq:Pt0}
P(t,\mathbf{r})=\langle \mathcal{T}V_L(t,\mathbf{r})e^{-\frac{i}{\hbar}\int_{-\infty}^tH_-(\tau)d\tau}\rangle,
\end{equation}
where $\langle...\rangle=\text{Tr}[\rho_0...]$ is understood where $\rho_0$ is the initial field plus matter density operator, $\mathcal{T}$ is time ordering operator.
  
The linear response is obtained by calculating the signal to second order in the coupling to the classical field. QED corrections to the linear response are obtained to fourth order in field matter interactions (two with the classical modes and two with vacuum modes). The transmitted classical field scales as $\sim|E(\omega)|^2$, which contributes to the linear response. The latter  correction to the linear response is phase independent, and cannot be manipulated by coherent control schemes. The lowest order contribution to the nonlinear response  that  contains phase information of the incoming pulse and has non additive contributions that may reveal collective resonances, involves six field matter interactions (four -  with classical broadband pulse and two - with quantum vacuum modes that mediate the interaction between atoms.)

We assume that  the system is initially in the ground state $g$. The relevant  diagrams responsible for collective effects when the last emission occurs from atom $A$ are shown in Fig. \ref{fig:6g}. Similar set of diagrams can be obtained when the last emission is with atom $B$. The total signal is given by the sum of the pathways corresponding to each diagram: $S_{A}(\omega)=\sum_iS_{Ai}(\omega)$, and can be read off the diagrams of Fig. \ref{fig:6g} (see Appendix \ref{app:6}). 

The classical response function is given by the diagrams in Fig. \ref{fig:6g}. These result in Eqs. (\ref{eq:S6100}) - (\ref{eq:S690}) which use normally ordered field operators. The field correlation function of normally ordered operators when the field is in a coherent state, which is the closest to the classical, may be factorized into a product of field amplitudes. Terms where the field operators are not-normally ordered exist in several pathways. They can be brought into a normally ordered form by making use of the commutation relations (\ref{eq:com}) - (\ref{eq:com1}). These apply to the quantum modes of the radiation field (wavy lines in Fig. \ref{fig:6g}) that are initially are in vacuum state.

The total signal including the diagrams where the last emission is with atom $B$ is

%\begin{widetext}
\begin{align}\label{eq:S6f}
S(\omega)&=\mathcal{I}\frac{N_{AB}|\mu_A|^2|\mu_B|^2}{2\pi\hbar^6}\int_{-\infty}^{\infty}\frac{d\omega_1}{2\pi}\frac{d\omega_2}{2\pi}\notag\\
&E^{*}(\omega)E^{*}(\omega_1)E(\omega+\omega_1-\omega_2)E(\omega_2)\notag\\
&\times\chi^{(3)}_{QED}(-\omega,-\omega_1,\omega+\omega_1-\omega_2,\omega_2)], 
\end{align}
%\end{widetext}
where $\mathcal{I}$ denotes the imaginary part. For reasons that will become clear later we partition $\chi_{QED}^{(3)}$ into two group of terms $\chi^{(3)}_{QED}=\chi^{(3)}_I+\chi^{(3)}_{II}$. Both can be read from the diagrams in Fig. \ref{fig:6g} and are given in Appendix \ref{app:6}. 

%A useful way of representing the susceptibility is through a superoperator notation. In particular the semiclassical contribution reads
\begin{align}\label{eq:csc0}
&\chi^{(3)I}=\sum_{j=1}^3\chi^{(3)I}_{jLLLL}(-\omega,-\omega_1,\omega+\omega_1-\omega_2,\omega_2)\notag\\
&+\sum_{j=4,5,7}[\chi^{(3)I}_{jLLLR}(-\omega,-\omega_1,\omega+\omega_1-\omega_2,\omega_2)+\chi^{(3)I}_{jL\leftrightarrow R}],
\end{align}
%whereas the QED correction yields
%\begin{align}
%\chi_{SC}^{(3)}=&-2\chi^{(5)}_{LLLLL-}(\omega,\omega_1,\omega+\omega_1-\omega_2,\omega_2,\omega_2,\omega_2)\notag\\
%&+2\chi^{(5)}_{LLLLRL}(\omega,\omega_1,\omega+\omega_1-\omega_2,\omega_2,\omega_2,\omega_2)\notag\\
%&+2\chi^{(5)}_{LLLRLL}(\omega,\omega_1,\omega+\omega_1-\omega_2,\omega_2,\omega_2,\omega_2)\notag\\
%&+2\chi^{(5)}_{LLRLLL}(\omega,\omega_1,\omega+\omega_1-\omega_2,\omega_2,\omega_2,\omega_2)\notag\\
%&+2\chi^{(5)}_{LRLLLL}(\omega,\omega_1,\omega+\omega_1-\omega_2,\omega_2,\omega_2,\omega_2).
%\end{align}
%It is interesting that semiclassical $\chi^{(3)}$ in superoperator notation can be recast via a classical response function $\chi_{+---}^{(3)}$ (neglecting fluorescence contributions). Furthermore the QED contribution via $\chi^{(3)}_Q$ is actually the $\chi^{(5)}$ with additional pairwise energy conservation for the pairs of the pulses:
\begin{align}\label{eq:cq0}
&\chi^{(3)II}=\int\frac{d\omega'}{2\pi}[\chi^{(5)II}_{1LLLLLL}(-\omega,-\omega_1,\omega',\omega+\omega_1-\omega_2,-\omega',\omega_2)\notag\\
&+\chi^{(5)II}_{2LLLLLL}(-\omega,\omega',-\omega_1,-\omega',\omega+\omega_1-\omega_2,\omega_2)\notag\\
&+\chi^{(5)II}_{4LLLLLR}(-\omega,-\omega_1,\omega',\omega+\omega_1-\omega_2,-\omega',\omega_2)+\chi^{(5)II}_{4L\leftrightarrow R}\notag\\
&+\chi^{(5)II}_{6LLLLRR}(-\omega,\omega',-\omega_1,-\omega',\omega+\omega_1-\omega_2,\omega_2)+\chi^{(5)II}_{6L\leftrightarrow R}],
\end{align}
where the numerical subscript corresponds to the diagrams in Fig. \ref{fig:6g} and various permutations of ``left'' and ``right'' interactions for diagrams 4 and 6 are included. It follows from the expressions given in Appendix \ref{app:6} that $\chi_{QED}^{(3)}$ contains a TPA collective resonance which contains Green's function $G_{ab}^{(+)}(\omega+\omega_1)=i/[\omega+\omega_1-\omega_{+}+i\gamma_{ab}]$ with $\omega_{+}=\omega_a+\omega_b$ and $\gamma_{ab}=\gamma_a+\gamma_b$.ÊHere $\gamma_{\alpha}^{-1}$, $\alpha=a,b$ represents the lifetime of state $\alpha$ that is ultimately related to the coupling constant (\ref{eq:Lab}) obtained from QME treatment: $\gamma_{\alpha}=\mathcal{L}_{\alpha\alpha}(\omega)$ \cite{Aga74}. The collective resonances $\omega+\omega_1=\omega_{+}+i\gamma_{ab}$ generally wash out by the $\omega_1$ integration. However we shall demonstrate how to these TPA resonances as well as collective Raman resonances can be recovered by pulse shaping.

\section{Detecting collective resonances by spectroscopy with shaped pulses}

  \begin{figure}[t]
\begin{center}
\includegraphics[trim=0cm 0cm 0cm 0cm,angle=0, width=0.45\textwidth]{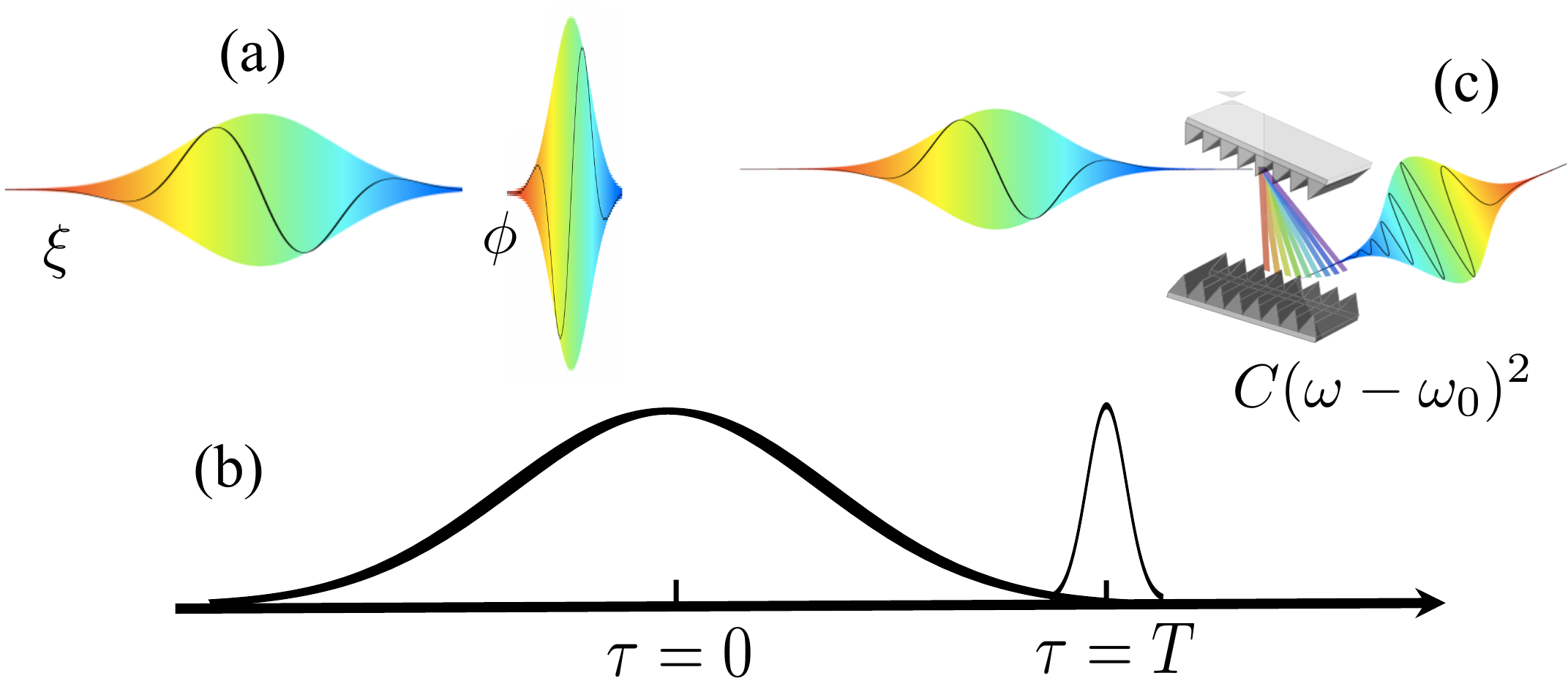}
\end{center}
\caption{(Color online) Pulse shaping for narrowband (picosecond) pulse with phase $\xi$ and broadband (femtosecond) pulse with phase $\phi(\omega)$. Constant phase $\phi$ - (a), linear phase (time delay $\phi(\omega)=\omega T$) - (b) and quadratic phase (linearly chirped $\phi(\omega)=C(\omega-\omega')^2$) - (c)}
\label{fig:ps}
\end{figure}

We assume an incoming classical pulse consisting of a long (picosecond) and broadband (femtosecond) pulses (see Fig. \ref{fig:ps}). The electric field reads
\begin{align}\label{eq:Ew}
E(\omega)=2\pi\mathcal{E}_1e^{i\xi}\delta(\omega-\omega_p)+2\pi\mathcal{E}_2e^{i\phi(\omega)}.
\end{align}
We shall use the amplitudes $\mathcal{E}_1$, $\mathcal{E}_2$ and the phases $\xi$ and $\phi(\omega)$ of these two fields as control parameters. The signal (\ref{eq:S6f}) depends on the following product of field amplitudes
\begin{widetext}
\begin{align}\label{eq:Ecor}
\frac{1}{(2\pi)^2} E^{*}(\omega)&E^{*}(\omega_1)E(\omega+\omega_1-\omega_2)E(\omega_2)\notag\\
=\mathcal{E}_2^2\mathcal{E}_1^2&\{\delta(\omega_1-\omega_p)[\delta(\omega_2-\omega)+\delta(\omega_2-\omega_p)]+\delta(\omega+\omega_1-2\omega_p)\delta(\omega_2-\omega)e^{i[2\xi-\phi(\omega)-\phi(\omega_1)]}\}\notag\\
+\mathcal{E}_2^3\mathcal{E}_1&\{\delta(\omega_1-\omega_p)e^{i[\phi(\omega+\omega_p-\omega_2)+\phi(\omega_2)-\phi(\omega)-\xi]}+\delta(\omega_2-\omega_p)e^{i[\phi(\omega+\omega_1-\omega_p)+\xi-\phi(\omega)-\phi(\omega_1)]}\notag\\
&+\delta(\omega+\omega_1-\omega_2-\omega_p)e^{i[\phi(\omega+\omega_1-\omega_p)+\xi-\phi(\omega)-\phi(\omega_1)]}\},
\end{align}
\end{widetext}
where the last interaction that results in the last emission occurs with a broadband field $\mathcal{E}_2$ and we neglect  $\sim\mathcal{E}_2^4$ term which contains no collective resonances in (\ref{eq:S6f}). We shall show that the $\sim\mathcal{E}_2^3\mathcal{E}_1$ terms contain interesting phase information.

We hold the amplitude and the phase of the narrowband pulse fixed and calculate the transmission of the broadband pulse while varying its parameters. Contour integration is used to evaluate the frequency integrations  in (\ref{eq:S6f}). We shall expand the phase $\phi(\omega)$ is a Taylor series in the vicinity of a reference frequency $\omega'$
\begin{align}
\phi(\omega, \{C_n\})=\sum_nC_n\cdot(\omega-\omega')^n.
\end{align}
$C_0$, $C_1$ and $C_2$ represent a constant phase. pulse delay and chirping, respectively. The sign of $C_n$ defines the direction of the contour in complex plane for evaluation of the residues in the frequency integrations. Assuming short delay, or small chirp rate the signal (\ref{eq:S6f}) becomes
\begin{widetext}
\begin{align}\label{eq:S6s}
S(\omega,\omega_p)=S_{I}(\omega,\omega_p)+S_{II}(\omega,\omega_p),
\end{align}
\begin{align}\label{eq:S6sc}
&S_{I}(\omega,\omega_p)=\mathcal{I}\frac{iN|\mu_A|^2|\mu_B|^2}{\hbar^6}(\mathcal{E}_2^3\mathcal{E}_1[G_{ab}^{(+)}(\omega +\omega_p)A_1(\omega,\omega_p)+G_{ab}^{(+)2}(\omega +\omega_p)A_2(\omega,\omega_p)\notag\\
+\sum_{\alpha,\beta,\delta}\{&G_{\beta\alpha}^{(-)}(\omega-\omega_p)A_3^{(\alpha)}(\omega,\omega_p)+G_{\beta\alpha}^{(-)}(\omega-\omega_p)G_{\delta\alpha}^{(-)}(\omega-\omega_p)A_4^{(\alpha\beta\delta)}(\omega,\omega_p)+G_{\beta\alpha}^{(-)\dagger}(\omega_p-\omega)G_{\delta\alpha}^{(-)\dagger}(\omega_p-\omega)A_5^{(\alpha\beta\delta)}(\omega,\omega_p)\}]\notag\\
+\mathcal{E}_2^2\mathcal{E}_1^2[&G_{ab}^{(+)}(\omega +\omega_p)A_6(\omega,\omega_p)+G_{ab}^{(+)}(2\omega_p)A_7(\omega,\omega_p)+G_{ab}^{(+)2}(\omega +\omega_p)A_8(\omega,\omega_p)+G_{ab}^{(+)2}(2\omega_p)A_9(\omega,\omega_p)])
\end{align}
\begin{align}\label{eq:S6q}
&S_{II}(\omega,\omega_p)=\mathcal{I}\frac{iN_{AB}|\mu_A|^2|\mu_B|^2}{\hbar^6}(\mathcal{E}_2^3\mathcal{E}_1[G_{ab}^{(+)}(\omega +\omega_p)B_1(\omega,\omega_p)+G_{ab}^{(+)}(\omega +\omega_p)\notag\\
\times [&G_{ab}^{(+)}(\omega+\omega_p)B_2(\omega,\omega_p)+G_{aa}^{(+)}(\omega+\omega_p)B_3(\omega,\omega_p)+G_{bb}^{(+)}(\omega+\omega_p)B_4(\omega,\omega_p)]\notag\\
+\sum_{\alpha,\beta}\{&G_{\beta\alpha}^{(-)}(\omega-\omega_p)B_5^{(\alpha\beta)}(\omega,\omega_p)+G_{\beta\alpha}^{(-)}(\omega_p-\omega)B_6^{(\alpha\beta)}(\omega,\omega_p)+G_{\beta\alpha}^{(-)\dagger}(\omega_p-\omega)B_7^{(\alpha\beta)}(\omega,\omega_p)\}]\notag\\
+\mathcal{E}_2^2\mathcal{E}_1^2[&G_{ab}^{(+)}(\omega +\omega_p)B_8(\omega,\omega_p)+G_{ab}^{(+)}(2\omega_p)B_9(\omega,\omega_p)]).
\end{align}
\end{widetext}
The parameters $A_1$ - $A_9$ and $B_1$ - $B_9$ are listed in Appendix \ref{app:par} and $N$ is the number of $A/B$ pairs. Here the collective Raman Green's function $G_{\alpha\beta}^{(-)}(\omega)=i/[\omega-\omega_{\alpha\beta-}+i\gamma_{\alpha\beta}]$ and collective TPA Green's function $G_{\alpha\beta}^{(+)}(\omega)=i/[\omega-\omega_{\alpha\beta+}+i\gamma_{\alpha\beta}]$ with $\omega_{\alpha\beta\pm}=\omega_{\alpha}\pm\omega_{\beta}$, $\gamma_{\alpha\beta}=\gamma_{\alpha}+\gamma_{\beta}$, $\alpha,\beta=a,b$.

\begin{figure*}[t]
\begin{center}
\includegraphics[trim=0cm 1cm 0cm 0cm,angle=0, width=0.95\textwidth]{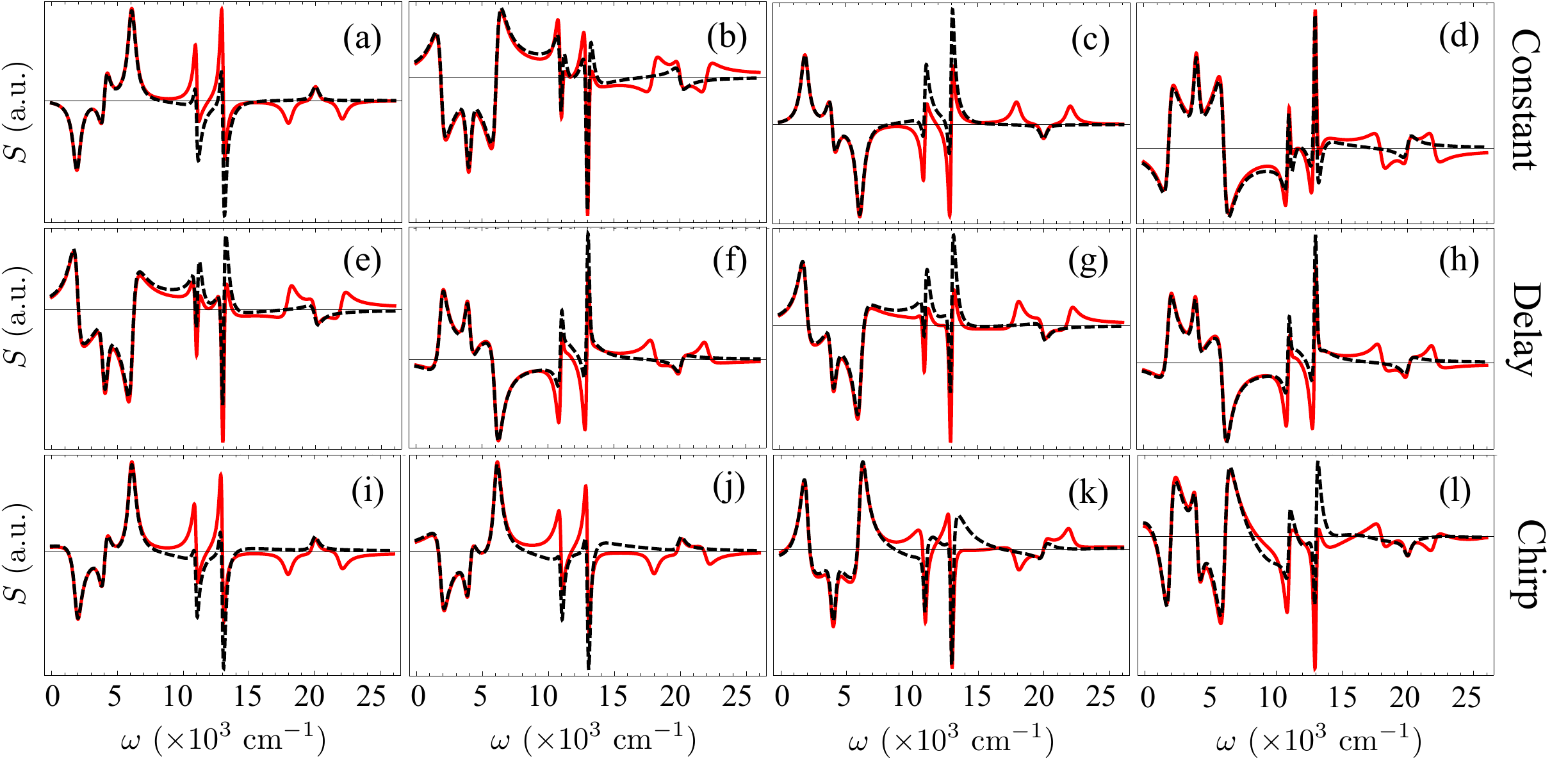}
\end{center}
\caption{(Color online) Top row: frequency dispersed transmission signal for pairs of atoms. The constant phases are $\Delta\phi=0$, $\pi/2$, $\pi$ and $3\pi/2$ for (a) - (d), respectively. Red solid line gives the full signal (\ref{eq:S6s}). Black dashed line represents Eq. (\ref{eq:S6sc}). Middle row: variable delay $T=17$ fs - (e), 33 fs - (f), 330 fs - (g), 3.3 ps - (h). Bottom row: linear chirp $C_2=5\cdot10^{-9}$ cm$^{-2}$ - (i), $10^{-8}$ cm$^{-2}$ - (j), $2.5\cdot10^{-8}$ cm$^{-2}$ - (k), and $5\cdot10^{-8}$ cm$^{-2}$ - (l).}
\label{fig:1D}
\end{figure*} 

The signals (\ref{eq:S6sc}) - (\ref{eq:S6q}) contain  the TPA Green's function $G_{ab}^{(+)}$ as well as Raman type collective resonances governed by Green's function $G_{\beta\alpha}^{(-)}$. The latter are of two types - $\alpha=\beta$ elastic (Rayleigh) scattering and  $\alpha\neq\beta$ Raman. The Rayleigh resonance contains a factor of 2 compare to Raman contribution due to permutations between $\alpha\leftrightarrow\beta$. This causes  $N^2$ vs $N$ scaling of the signal, respectively. 

The narrow and broadband field amplitudes allow for additional control over the resonance features. If the broadband pulse is strong, the signal  (\ref{eq:S6sc}) generated by a pair of different atoms $A/B$ shows only one type of TPA resonance $\omega+\omega_p=\omega_a+\omega_b$ whereas (\ref{eq:S6q}) has two additional additional $\omega+\omega_p=2\omega_a$, $2\omega_b$ (see Appendix \ref{app:par}). The latter resonances are missing if multiple interactions occur within the single atom and therefore constitute collective nature. Clearly these type of resonances will appear in  the signal (\ref{eq:S6sc}) for a pair of atoms of the same type, $A/A$ or $B/B$. However, in an arbitrary sample composed by several species depending on the density of the sample as well as the dipole moments $\mu_A$ vs $\mu_B$  it is possible to obtain the couplings between atoms of different types. In certain parameter regime, for example for a gas of two types of atoms signals $S_I$ and $S_{II}$ predict different resonances.

\section{Simulations}

We now compare the relative strength of all collective resonances and show how they can be controlled by the nonlinear phase $\phi(\omega)$. Consider a system of two atoms $A/B$ with transition frequencies $\omega_a=13000$ cm$^{-1}$, $\omega_b=11000$ cm$^{-1}$ and linewidth $\gamma_a=\gamma_b=200$ cm$^{-1}$. In Fig. \ref{fig:1D} we depict the signal $S(\omega,\omega_p)$ vs the broadband frequency $\omega$ for a fixed off-resonant $\omega_p=4000$ cm$^{-1}$ and $\mu_B\simeq 0.99\mu_A$. In this section we only discuss the full signal given by Eqs. (\ref{eq:S6s}) (red solid line). The black dashed line which shows the $S_I$ contribution (Eq. (\ref{eq:S6sc})) will be discussed in Section V.

It is apparent that only $\mathcal{E}_2^3\mathcal{E}_1$ terms in Eqs. (\ref{eq:S6sc}) - (\ref{eq:S6q})  contain Raman-type resonance whereas the $\mathcal{E}_2^2\mathcal{E}_1^2$ terms yield TPA resonances. The TPA resonance is weaker than the Raman and single-photon resonance in most of the parameter regimes. One can probe the $\mathcal{E}_2^3\mathcal{E}_1$ and $\mathcal{E}_2^2\mathcal{E}_1^2$ terms separately due to the different intensity dependence.

\begin{figure*}[t]
\begin{center}
\includegraphics[trim=0cm 1cm 0cm -0.5cm,angle=0, width=0.95\textwidth]{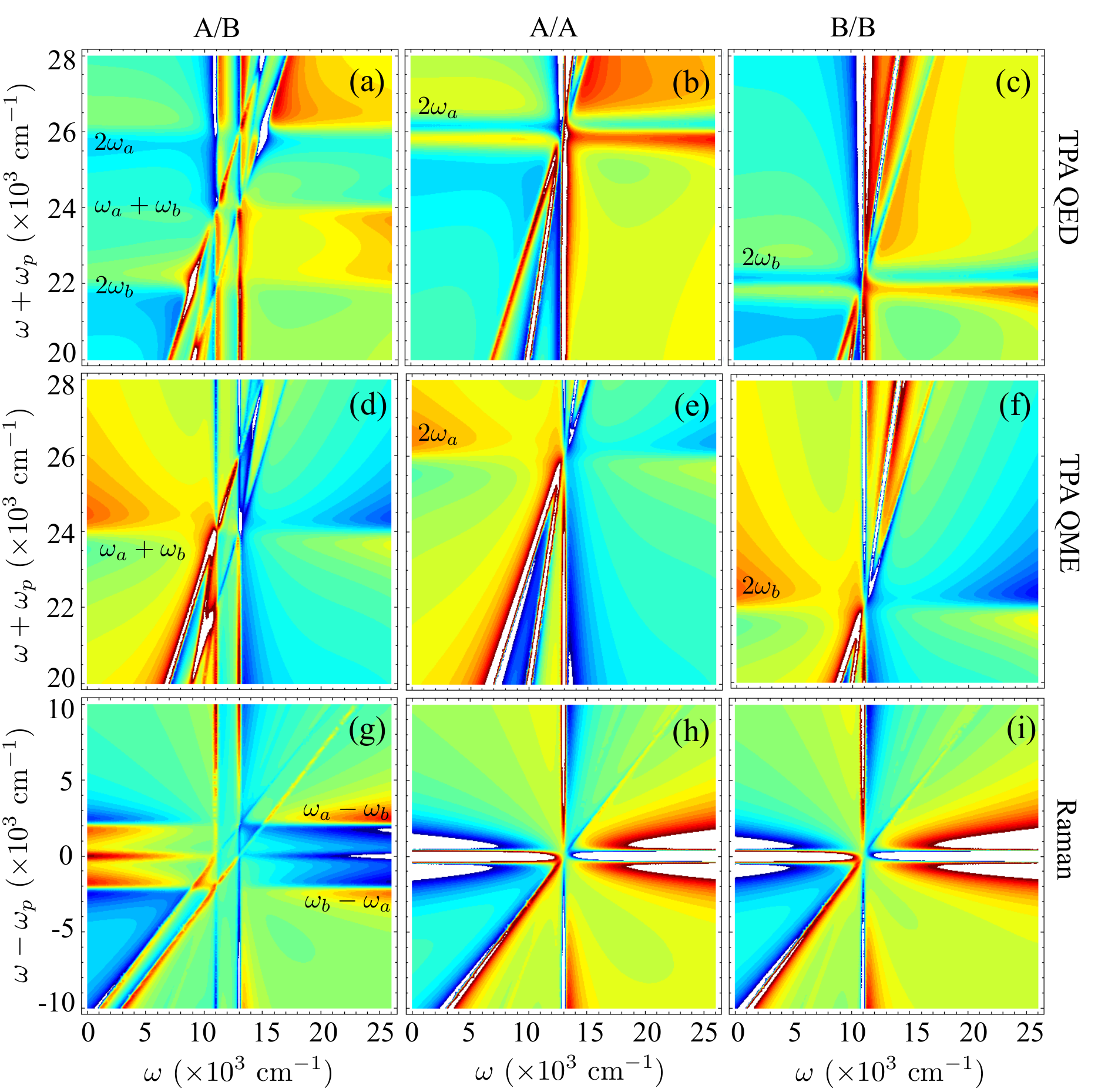}
\end{center}
\caption{(Color online) Frequency dispersed residue signal (\ref{eq:Sr}) with linear chirped broadband pulse with chirp rate $C_2=5\cdot10^{-9}$ cm$^{-2}$. Left column: system $A/B$, middle: $A/A$, right: $B/B$. Top row: result of Eq. (\ref{eq:S6s}), middle: result of Eq. (\ref{eq:S6sc})  correspond to TPA resonances via $S_r(\omega+\omega_p,\omega)$, bottom: Raman resonances via $S_r(\omega-\omega_p,\omega)$ using full signal (\ref{eq:S6s}).}
\label{fig:2D}
\end{figure*}  

In the following simulations we focus on the $\mathcal{E}_2^3\mathcal{E}_1$ that contain both Raman and TPA collective resonances.
We consider three models for the phase $\phi(\omega)$. Model $i$ - a constant phase: $\phi(\omega)=\xi+\Delta\phi$; a linear phase ($ii$) \cite{Kon12}: $\phi(\omega)=\omega T$ induces a delay $T$ of the broadband pulse relative to the narrowband; and finally quadratic phase ($iii$)  $\phi(\omega)=C_2(\omega-\omega')^2$ which represents linear chirp \cite{Kon121} with reference frequency $\omega'=(\omega_a+\omega_b)/2$.

We start with model $i$. Fig. \ref{fig:1D}a shows that for a fixed off-resonant narrowband frequency $\omega_p=4000$ cm$^{-1}$ the spectra has two Raman  $\omega=2000$, $6000$ and one Rayleigh peaks $\omega\sim4000$ cm$^{-1}$, two single-photon resonances at $\omega=\omega_a,\omega_b$ and two-photon resonance $\omega=\omega_a+\omega_b-\omega_p\sim 20000$ cm$^{-1}$. The signal (\ref{eq:S6s}) contains two additional TPA peaks at $\omega=2\omega_a-\omega_p=22000$ cm$^{-1}$ and $\omega=2\omega_b-\omega_p=18000$ cm$^{-1}$. We next turn to the TPA resonances. If $\Delta\phi_0=0,\pi$ (Fig. \ref{fig:1D}a,c ) peak at $\omega_a+\omega_b$, which corresponds to emission and have a dip at $2\omega_a$ and $2\omega_b$ corresponding to absorption. For $\Delta\phi_0=\pi/2,3\pi/2$ (Fig. \ref{fig:1D}b,d) the situation is different. It corresponds to a destructive quantum interference of absorption/emission pathways corresponding to two atoms (e.g. Fano interference) and all three TPA peaks become asymmetric. 

Model $ii$ is shown in panels e-h. The interplay between destructive interference (asymmetric), absorption (dip) and emission (peak) for the TPA resonances is less susceptible to the delay than model $i$ for the phase shift. All three TPA peaks show quantum interference for the entire range of the delays from 17 fs to 3.3 ps shown in  Figs. \ref{fig:1D}e-h.  

The resonance pattern for model $iii$ is more complex. For small positive chirp rate $C_2=5\cdot 10^{-9}-10^{-8}$ cm$^{-2}$ - Fig. \ref{fig:1D}i,j all three TPA peaks are symmetric where  $2\omega_a$, $2\omega_b$  correspond to absorption (dip) whereas $\omega_a+\omega_b$ has a peak (emission). For moderate chirp rate $C_2=2.5\cdot 10^{-8}$ cm$^{-2}$ - Fig. \ref{fig:1D}k the $\omega_a+\omega_b$  peak becomes asymmetric which corresponds to the regime of (destructive interference) whereas two symmetric peaks at $2\omega_a$ and $2\omega_b$ now have different sign corresponding to emission (peak) of one and absorption (dip) of another. For  the larger chirp rate $C_2=5\cdot 10^{-8}$ cm$^{-2}$ (Fig. \ref{fig:1D}k) the collective resonances are slightly less pronounced compared to the single photon peaks, whereas for negative chirp $C_2=-5\cdot 10^{-8}$ cm$^{-2}$  - Fig. \ref{fig:1D}l the situation becomes the opposite: two peaks $2\omega_a$ and $2\omega_b$ are asymmetric (interference) whereas $\omega_a+\omega_b$ corresponds to absorption (dip).

To better distinguish between various collective and single photon resonances we display a 2D signal vs the broadband $\omega$ and the narrowband $\omega_p$ frequencies. Chirping ($iii$) allows to eliminate the background by looking at the residue signal shown in Fig. \ref{fig:2D} defined as the difference of two measurements with opposite sign of chirp 
\begin{equation}\label{eq:Sr}
S_r(\omega,\omega_p)\equiv S(\omega,\omega_p,C_2)-S(\omega,\omega_p,-C_2).
\end{equation}

Fig. \ref{fig:2D}a shows the $A/B$ system. It contains two types of single photon resonances shown by vertical lines due to single photon resonance with broadband field at $\omega=\omega_a$ and $\omega_b$ and 45 degrees inclined lines corresponding to the single photon resonance with narrow band field at $\omega_p=\omega_a$ and $\omega_b$. In addition, we observe three collective TPA peaks depicted by horizontal lines at $\omega+\omega_p= \omega_a+\omega_b=24000$ cm$^{-1}$, $\omega+\omega_p=2\omega_a=26000$ cm$^{-1}$ and $\omega+\omega_p=2\omega_b=22000$ cm$^{-1}$ as predicted by Eq. (\ref{eq:S6s}). For the systems $A/A$ and $B/B$ shown in Fig. \ref{fig:2D}b,c, repsectively, the corresponding collective resonance is given by a single TPA resonance at $\omega+\omega_p=2\omega_a$ and $\omega+\omega_p=2\omega_b$, respectively. 

Calculation using partial signal (\ref{eq:S6sc}) results in a single TPA resonance for all three types of system: $\omega_\alpha+\omega_\beta$ for $\alpha+\beta$, $\alpha,\beta=A,B$, which is illustrated by Fig. \ref{fig:2D}d-f. The latter arises from diagram 4 in Fig. \ref{fig:6g} and corresponds to the following situation. Initial excitation by the incoming pulse that acts on both bra and ket brings the system to the nonstationary density matrix which is then radiates a spontaneous photon leaving the system in the excited-to-ground state coherence. After the second interaction with incoming pulse which promotes the system to a single excited state the spontaneous photon emitted by the first atom is finally absorbed by the second atom that forces the the two-atom system to the double-to-single excited state coherence. It then undergoes a stimulated emission via the interaction with the incoming pulse for the fourth time and the system ends up in the single excited state population state. In the following we will mostly focus on TPA type collective resonances.

The bottom row of Fig. \ref{fig:2D} shows collective Raman resonances accessible by a signal (\ref{eq:S6s}). Fig. \ref{fig:2D}g for $A/B$ contains an elastic (Rayleigh) resonance at $\omega=\omega_p$  and two Raman resonances $\omega-\omega_p=\pm\omega_a\mp\omega_b$. The $A/A$ and $B/B$ signal only shows a Rayleigh peak (the side peaks appear due to oscillation of the nonlinear phase in the residue signal (\ref{eq:Sr})).

\section{Collective resonances predicted by the QME}

The SC approach describes the coupling mediated by exchange of photons by a Quantum Master Equation (QME) for the matter density operator
\begin{align}\label{eq:ME}
&\dot{\rho}=-i\sum_{\alpha}[\omega_\alpha^{(0)}\sigma_\alpha^{(z)},\rho]-i\sum_{\alpha\neq\beta}\mathcal{L}_{\alpha\beta}[V_\alpha^{\dagger}V_\beta,\rho]\notag\\
&-\sum_{\alpha,\beta}\gamma_{\alpha\beta}(V_\alpha^{\dagger}V_\beta\rho-2V_\beta\rho V_\alpha^{\dagger}+\rho V_\alpha^{\dagger}V_\beta)-i[H_{int}^{(c)},\rho],
\end{align}
where $\omega_\alpha^{(0)}$ is renormalized transition frequency, $\mathcal{L}_{\alpha\beta}$ is the dipole-dipole interaction due to interaction with the common quantum mode, $\gamma_{\alpha\beta}$ is a cooperative emission rate. The last term represents the interaction with classical field modes. 

Some quantum pathways for the signal (\ref{eq:S0}) can be obtained directly from the QME (\ref{eq:ME}) as can be deduced from the corresponding diagrams shown in Fig. \ref{fig:6g}. If two consecutive interactions occur with quantum modes, then the signal can be obtained in the lower order $\chi^{(3)}$ theory rather than $\chi^{(5)}$ by introducing an effective interatomic couplings that originate from emission and reabsorption of the photon by a single excited state of the system through the ground state (diagrams 1, 2, 4, and 7 in Fig. \ref{fig:6g}) 
\begin{align}\label{eq:Lc}
&\mathcal{L}_{\alpha\beta}(\omega)=\frac{1}{\hbar^2}\int_{-\infty}^{\infty}\frac{d\omega'}{2\pi}\mu_\alpha^{(l)*}\mu_\beta^{(m)}\mathcal{D}_{\alpha\beta}^{(l,m)}(\omega')G_g(\omega-\omega'),
\end{align}
where summation is assumed for repeating indices. Similar two-photon coupling for emission and reabsorption by a two-photon state of the system through single photon state (diagrams 3 and 5 in Fig. \ref{fig:6g}) gives
\begin{align}\label{eq:Ls}
\mathcal{L}_s(\omega+\omega_1)=\frac{1}{\hbar^2}\sum_{\alpha}\int_{-\infty}^{\infty}\frac{d\omega'}{2\pi}\mathcal{L}_{\alpha\alpha}(\omega')G_{\bar{\alpha}}(\omega+\omega_1-\omega'),
\end{align}
where $G_\alpha=i/(\omega-\omega_\alpha+i\gamma_{\alpha})$ and $\bar{\alpha}=a$ if $\alpha=b$ and $\bar{\alpha}=b$ if $\alpha=a$.

\begin{figure}[t]
\begin{center}
\includegraphics[trim=0cm 1cm 0cm -0.5cm,angle=0, width=0.45\textwidth]{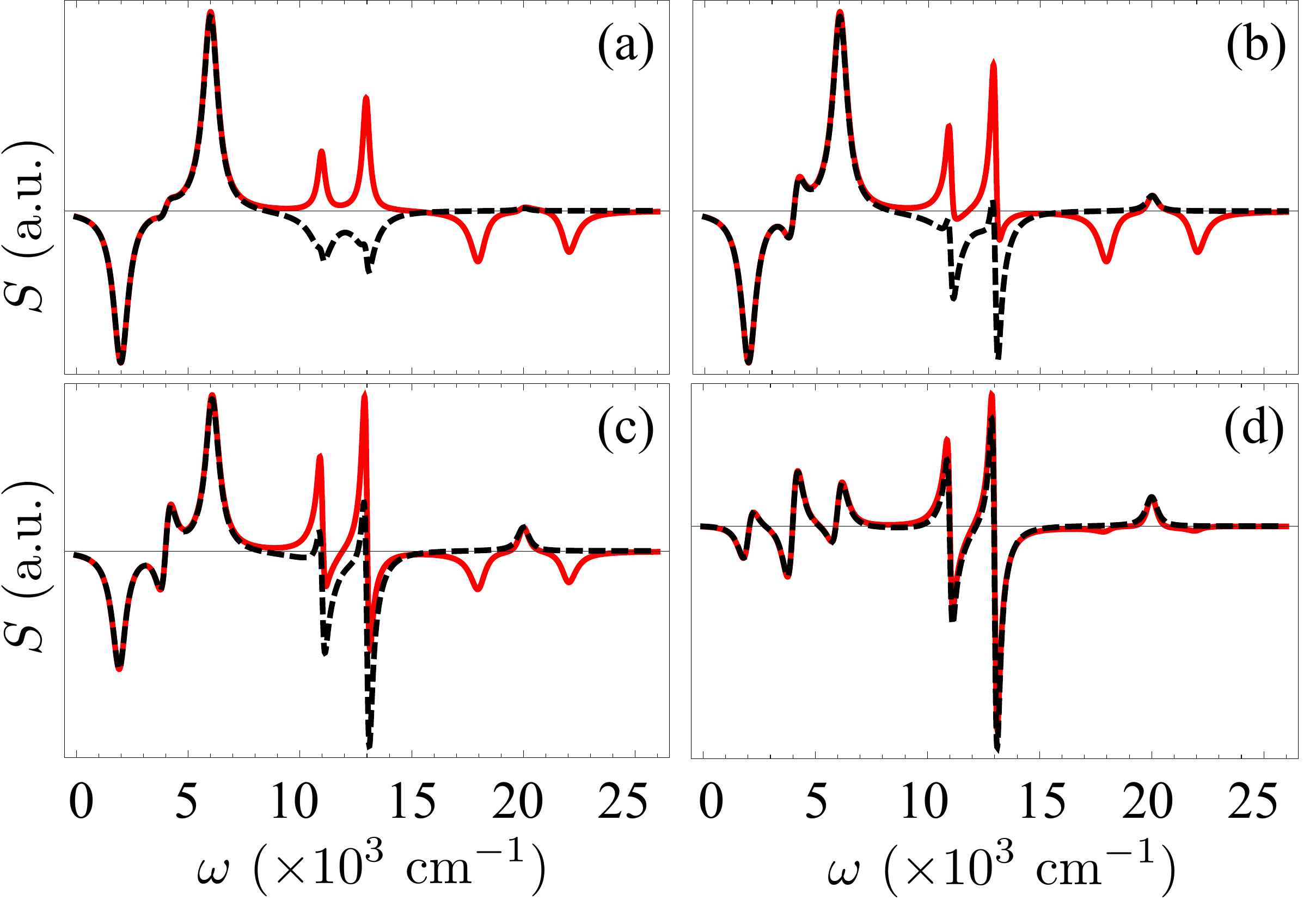}
\end{center}
\caption{(Color online) Frequency dispersed transmission from two atoms with a shaped pulse with zero phase $\xi=0$, $\phi(\omega)=0$ for different distances between atoms $r_{\alpha\beta}/\lambda_A$ 0.001 - (a), 0.005 - (b), 0.01 - (c) and 0.1 - (d).}
\label{fig:1Dp}
\end{figure}    

Eqs. (\ref{eq:Lc}) - (\ref{eq:Ls}) reveal that diagrams 3, 5, and 7 of Fig. \ref{fig:6g} can be recast via master equation (\ref{eq:ME}). This is not the case for diagram 6, since two interactions with quantum modes have an additional interaction with classical filed in between. Using the same reasoning one can show that the remaining diagrams 1, 2, and 4 have both types of contributions - ones that can and ones that cannot be recast as an effective couplings. $\chi_{I}^{(3)}$ in Eq. (\ref{eq:csc0}) thus represents the QME contribution whereas $\chi_{II}^{(3)}$ in Eq. (\ref{eq:cq0}) requires the full QED description. 

To explain the limitations of the QME approach we first note that the signal strongly depends on the interatomic distance. We combine the density of radiation modes (\ref{eq:Dab}) and coupling in Eq. (\ref{eq:Lc}) using an identity \cite{Salam}
\begin{align}
&(-\nabla^2\delta_{\mu\nu}+\nabla_\mu\cdot\nabla_\nu)\frac{e^{ikR}}{R}=\notag\\
&\frac{1}{R^3}[(\delta_{\mu\nu}-3\hat{R}_\mu\hat{R}_\nu)(ikR-1)+(\delta_{\mu\nu}-\hat{R}_\mu\hat{R}_\nu)k^2R^2]e^{ikR}.
\end{align}
Assuming randomly orientated atoms: $\hat{R}_\mu\hat{R}_\nu=\frac{1}{3}\delta_{\mu\nu}$ we obtain $1/r_{\alpha\beta}$ distance dependence. Furthermore Eqs. (\ref{eq:S6sc}) - (\ref{eq:S6q}) contain coefficients $A_j$ and $B_j$, $j=1-9$ that depend on the distance between atoms via two types of couplings (see Appendix \ref{app:par}). The coupling (\ref{eq:Lc}) which gives the  cooperative decay rate \cite{Aga74}
\begin{align}\label{eq:Lab}
&\mathcal{L}_{\alpha\beta}(\omega)=\frac{\mu_\alpha^{*}\mu_\beta\omega^2}{3\pi\hbar\epsilon_0c^2 r_{\alpha\beta}}\sin[\omega r_{\alpha\beta}/c], \quad\mathcal{L}_{\alpha\alpha}(\omega)=\frac{|\mu_\alpha|^2\omega^3}{3\pi\hbar\epsilon_0c^3}.
\end{align}
This rate is typically small compare to the transition frequencies (weak coupling regime) $\mathcal{L}_{\alpha\beta}\ll \omega_{\alpha}$ and is finite at small distances due to the $\sin x/x$ factor. It enters the coefficients for most single-photon resonances and some collective Raman resonances. In addition there is a complex coupling (see Appendix \ref{app:par})
\begin{align}\label{eq:Mab}
\mathcal{M}_{\alpha\beta}(\omega)&=\frac{\mu_\alpha^{*}\mu_\beta\omega^2}{6\pi \hbar\epsilon_0c^2r_{\alpha\beta}}[i\cos(\omega r_{\alpha\beta}/c)+\sin(\omega r_{\alpha\beta}/c)],
\end{align}
where the first term corresponds to a dipole-dipole interaction and the second term is half of the cooperative spontaneous emission (superradiance) rate (\ref{eq:Lab}). Note that the dipole-dipole coupling grows rapidly $\sim r_{\alpha\beta}^{-3}$ at short distances. Our TPA resonances that depend on coefficients $B_3$ and $B_4$ in Eq. \ref{eq:S6q} are prominent at small atomic separation.

Fig. \ref{fig:1Dp} shows the variation of spectra with interatomic distance. For short distance compared to the wavelength it gives a  significant contribution showing new collective TPA resonances along with strong single-photon resonances. This varies at large distances according to Eq. (\ref{eq:S6q}). The Raman resonances behave similarly and become weaker with distance in both (\ref{eq:S6sc}) and (\ref{eq:S6q}).

Generally, the QED susceptibility (\ref{eq:cq0}) cannot be expressed via the effective coupling through the QME Eq. (\ref{eq:csc0}). However in the absence of a bath, setting the ground state frequency and linewidth to be zero $\omega_g=0$, $\gamma_g=0$, we have  $G_g(\omega)\simeq\delta(\omega)$. Thus, both susceptibilities (\ref{eq:csc0}) - (\ref{eq:cq0}) are governed by the small parameter which is related to the couplings $\mathcal{L}_{\alpha\beta},\mathcal{M}_{\alpha\beta}$ given by Eqs. (\ref{eq:Lab}) - (\ref{eq:Mab}), where$|\mathcal{L}_{\alpha\beta}(\omega)|,|\mathcal{M}_{\alpha\beta}(\omega)|\ll |\omega_\alpha-\omega_\beta|, \sigma$, where $\sigma$ is the bandwidth of the pulse envelopes. As we show below these couplings enter both QME and QED contributions in the different way. 

The magnitude of the QED correction is governed by a combined spectral bandwidth of matter and field degrees of freedom that enter the susceptibilities. This can be best understood in the joint field plus matter space. Due to the consecutive interactions with quantum modes in SC theory, the effective frequency range that enters SC susceptibility  (\ref{eq:csc1}) - (\ref{eq:csc7}) is governed by entire spectrum of quantum modes. On the other hand, due to the mixed time ordering of interactions with quantum and classical modes, the effective frequency range that enters the QED susceptibility (\ref{eq:cq1}) - (\ref{eq:cq6}) is limited by a combined classical pulse and matter bandwidth. To illustrate this we note that  Eqs. (\ref{eq:csc1}) and (\ref{eq:cq1}) show that for the same set of diagrams (1 and 4 in Fig. \ref{fig:6g}) the $\omega_2$ dependence enters Eqs. (\ref{eq:csc0}) and (\ref{eq:cq0}) via $G_\beta(\omega_2)$ and $G_\beta(\omega+\omega_1-\omega_2)$, $\beta=a,b$ respectively. Therefore  the frequency range of $\omega_2$ in the QED susceptibility is restricted compared to its SC counterpart. Another way to look at it is by noting that the collective resonance in (\ref{eq:csc1}) and (\ref{eq:cq14}) is given  by $G_{ab}^{(+)}(\omega+\omega_1)$. Because in the SC signal the integrations over $\omega_1$ and $\omega_2$ are uncoupled, if collective resonance exists and is not smeared by the pulse envelope it will enter through the same $G_{ab}^{(+)}(\omega+\omega_1)$. On the other hand, due to mixing of the frequency arguments in QED contribution, integration over $\omega_2$ may bring another collective resonance (e.g. $\omega_2=\omega_\alpha-i\gamma_\alpha$) that will now appear through the product of two Green's functions $G_{ab}^{(+)}(\omega+\omega_1)G_\beta(\omega+\omega_1-\omega_\alpha+i\gamma_\alpha)$ which gives rise to terms like $G_{ab}^{(+)2}(\omega+\omega_1)$, $G_{aa}^{(+)2}(\omega+\omega_1)$ and $G_{bb}^{(+)2}(\omega+\omega_1)$. In this case, the characteristic coupling accompanying such resonances will be $\tilde{\mathcal{M}}_{\alpha\beta}(\omega_\alpha-i\gamma_\alpha)$ given by Eq. (\ref{eq:Mab}) which grows rapidly at small distances. A full QED treatment contains fine details that are missed by the QME.

\section{Discussion}

The QME approach has had many successes and is widely used for calculating the third order response of a collection of two-level atoms. The QME is obtained by a second order expansion in the field-matter coupling strength, which is equivalent to introducing an effective interatomic coupling. One can further diagonalize the Hamiltonian and take into account the interatomic coupling to all orders. The perturbative QED treatment of the present paper shows, that in each order in field-matter coupling there are processes that are missed by the QME.

%The QED treatment results in new collective two photon absorption (TPA) resonances \cite{Ric11}. 

We have expressed the transmitted signal in terms of a four-point correlation function of the classical fields with an arbitrary number of quantum modes. We presented a QED calculation of the collective resonances to second order in the coupling to quantum vacuum modes. Radiative energy transfer between excited state populations \cite{Juz99} involves four interactions with the quantum modes (two with $A$ and two with $B$) and goes beyond the present theory. The QME which is based on the effective coupling stemming from two interactions with quantum mode can only describe certain type of collective resonances (Raman) but not TPA. The QED approach reveals resonant features stemming from nonconsecutive in time interactions with quantum modes  i.e. with classical field interaction in between. The magnitude of these contribution is governed by the dipole-dipole coupling and the combined spectral bandwidth of the relevant field and matter degrees of freedom. Generally, the QED contribution to the susceptibility that involves the sum over radiation modes over restricted frequency range is comparable to SC due to mixed time ordering of interactions. On the other hand the SC susceptibility contains consecutive interactions with quantum modes and thus requires a summation over the entire spectrum of these modes.  The use of pulse shaping (the combination of narrow - and broadband pulses) is crucial for observing these collective resonances. These resonances in the transmission of the shaped pulse can be best visualized by 2D plots vs the narrowband and broadband frequency. The former serves as a frequency reference  and the latter is dispersed by the detection. In addition, nonlinear phase shaping involving positive and negative chirp combination allows to eliminate the background and obtain a clear picture of the resonant features that include both collective and single photon resonances. The pulse Phase and amplitude may be used to manipulate the desired resonances.

The present approach is not restricted to classical states of the transmitted pulse and can be easily extended to different types of light, e.g. stochastic or entangled light. These will enter  the signal (\ref{eq:S6f}) via the four point correlation function of the incoming field \cite{Sch12}. Furthermore the formalism is not restricted to stimulated signals and can be applied to spontaneous signals as well. One of the potential applications may be to study the collective vs non collective features in aggregates with vibronic couplings \cite{Spa11} using photon counting signals \cite{Let10,Dor12}. The single molecule systems can also benefit from the proposed method \cite{Rez12}. 

\begin{acknowledgments}
The support of the National Science Foundation (Grant CHE-1058791), the Chemical Sciences, Geosciences, and Biosciences division, Office of Basic Energy Sciences, Office of Science, U.S. Department of Energy is gratefully acknowledged.  We also gratefully acknowledge the support of  of the National Institutes of health (Grant GM-59230). 
\end{acknowledgments}

\appendix

\begin{widetext}

\section{Signal contributions corresponding to Fig. 2}\label{app:6}

We read off the Liouville pathways  from the diagrams in Fig. \ref{fig:6g}

\begin{align}\label{eq:S6100}
S_{A1}(\omega)=-\mathcal{I}\frac{2i}{\hbar^6}&\int_{-\infty}^{\infty}dte^{i\omega t}\int_{-\infty}^td\tau_1\int_{-\infty}^{\tau_1}d\tau_2\int_{-\infty}^{\tau_2}d\tau_3\int_{-\infty}^{\tau_3}d\tau_4\int_{-\infty}^{\tau_4}d\tau_5\int d\mathbf{r}_1d\mathbf{r}_2d\mathbf{r}_3d\mathbf{r}_4d\mathbf{r}_5\notag\\
\times&\langle \mathcal{T}E_L^{\dagger}(\omega,\mathbf{r}_a)E_L^{\dagger}(\tau_1,\mathbf{r}_1)E_L(\tau_2,\mathbf{r}_2)E_L(\tau_3,\mathbf{r}_3)E_L^{\dagger}(\tau_4,\mathbf{r}_4)E_L(\tau_5,\mathbf{r}_5)\rangle\notag\\
\times&\langle \mathcal{T}V_L(t,\mathbf{r}_a)V_L(\tau_1,\mathbf{r}_1)V_L^{\dagger}(\tau_2,\mathbf{r}_2)V_L^{\dagger}(\tau_3,\mathbf{r}_3)V_L(\tau_4,\mathbf{r}_4)V_L^{\dagger}(\tau_5,\mathbf{r}_5)\rangle,
\end{align}
\begin{align}
S_{A2}(\omega)=-\mathcal{I}\frac{2i}{\hbar^6}&\int_{-\infty}^{\infty}dte^{i\omega t}\int_{-\infty}^td\tau_1\int_{-\infty}^{\tau_1}d\tau_2\int_{-\infty}^{\tau_2}d\tau_3\int_{-\infty}^{\tau_3}d\tau_4\int_{-\infty}^{\tau_4}d\tau_5\int d\mathbf{r}_1d\mathbf{r}_2d\mathbf{r}_3d\mathbf{r}_4d\mathbf{r}_5\notag\\
\times&\langle \mathcal{T}E_L^{\dagger}(\omega,\mathbf{r}_a)E_L(\tau_1,\mathbf{r}_1)E_L^{\dagger}(\tau_2,\mathbf{r}_2)E_L^{\dagger}(\tau_3,\mathbf{r}_3)E_L(\tau_4,\mathbf{r}_4)E_L(\tau_5,\mathbf{r}_5)\rangle\notag\\
\times&\langle \mathcal{T}V_L(t,\mathbf{r}_a)V_L^{\dagger}(\tau_1,\mathbf{r}_1)V_L(\tau_2,\mathbf{r}_2)V_L(\tau_3,\mathbf{r}_3)V_L^{\dagger}(\tau_4,\mathbf{r}_4)V_L^{\dagger}(\tau_5,\mathbf{r}_5)\rangle,
\end{align}
\begin{align}
S_{A3 }(\omega)=-\mathcal{I}\frac{2i}{\hbar^6}&\int_{-\infty}^{\infty}dte^{i\omega t}\int_{-\infty}^td\tau_1\int_{-\infty}^{\tau_1}d\tau_2\int_{-\infty}^{\tau_2}d\tau_3\int_{-\infty}^{\tau_3}d\tau_4\int_{-\infty}^{\tau_4}d\tau_5\int d\mathbf{r}_1d\mathbf{r}_2d\mathbf{r}_3d\mathbf{r}_4d\mathbf{r}_5\notag\\
\times&\langle\mathcal{T} E_L^{\dagger}(\omega,\mathbf{r}_a)E_L^{\dagger}(\tau_1,\mathbf{r}_1)E_L(\tau_2,\mathbf{r}_2)E_L^{\dagger}(\tau_3,\mathbf{r}_3)E_L(\tau_4,\mathbf{r}_4)E_L(\tau_5,\mathbf{r}_5)\rangle\notag\\
\times&\langle \mathcal{T}V_L(t)V_L(\tau_1,\mathbf{r}_1)V_L^{\dagger}(\tau_2,\mathbf{r}_2)V_L(\tau_3,\mathbf{r}_3)V_L^{\dagger}(\tau_4,\mathbf{r}_4)V_L^{\dagger}(\tau_5,\mathbf{r}_5)\rangle,
\end{align}
\begin{align}
S_{A4 }(\omega)=\mathcal{I}\frac{2i}{\hbar^6}&\int_{-\infty}^{\infty}dte^{i\omega t}\int_{-\infty}^td\tau_1\int_{-\infty}^{\tau_1}d\tau_2\int_{-\infty}^{\tau_2}d\tau_3\int_{-\infty}^{\tau_3}d\tau_4\int_{-\infty}^{t}d\tau_5\int d\mathbf{r}_1d\mathbf{r}_2d\mathbf{r}_3d\mathbf{r}_4d\mathbf{r}_5\notag\\
\times&\langle \mathcal{T}E_L^{\dagger}(\omega,\mathbf{r}_a)E_L(\tau_1,\mathbf{r}_1)E_L(\tau_2,\mathbf{r}_2)E_L^{\dagger}(\tau_3,\mathbf{r}_3)E_L(\tau_4,\mathbf{r}_4)E_R^{\dagger}(\tau_5,\mathbf{r}_5)\rangle\notag\\
\times&\langle \mathcal{T}V_L(t,\mathbf{r}_a)V_L^{\dagger}(\tau_1,\mathbf{r}_1)V_L^{\dagger}(\tau_2,\mathbf{r}_2)V_L(\tau_3,\mathbf{r}_3)V_L^{\dagger}(\tau_4,\mathbf{r}_4)V_R(\tau_5,\mathbf{r}_5)\rangle,
\end{align}
\begin{align}
S_{A5 }(\omega)=\mathcal{I}\frac{2i}{\hbar^6}&\int_{-\infty}^{\infty}dte^{i\omega t}\int_{-\infty}^td\tau_1\int_{-\infty}^{\tau_1}d\tau_2\int_{-\infty}^{\tau_2}d\tau_3\int_{-\infty}^{\tau_3}d\tau_4\int_{-\infty}^{t}d\tau_5\int d\mathbf{r}_1d\mathbf{r}_2d\mathbf{r}_3d\mathbf{r}_4d\mathbf{r}_5\notag\\
\times&\langle \mathcal{T}E_L^{\dagger}(\omega,\mathbf{r}_a)E_L(\tau_1,\mathbf{r}_1)E_L^{\dagger}(\tau_2,\mathbf{r}_2)E_L(\tau_3,\mathbf{r}_3)E_L(\tau_4,\mathbf{r}_4)E_R^{\dagger}(\tau_5,\mathbf{r}_5)\rangle\notag\\
\times&\langle \mathcal{T}V_L(t,\mathbf{r}_a)V_L^{\dagger}(\tau_1,\mathbf{r}_1)V_L(\tau_2,\mathbf{r}_2)V_L^{\dagger}(\tau_3,\mathbf{r}_3)V_L^{\dagger}(\tau_4,\mathbf{r}_4)V_R(\tau_5,\mathbf{r}_5)\rangle,
\end{align}
\begin{align}
S_{A6 }(\omega)=-\mathcal{I}\frac{2i}{\hbar^6}&\int_{-\infty}^{\infty}dte^{i\omega t}\int_{-\infty}^td\tau_1\int_{-\infty}^{\tau_1}d\tau_2\int_{-\infty}^{\tau_2}d\tau_3\int_{-\infty}^{t}d\tau_4\int_{-\infty}^{\tau_4}d\tau_5\int d\mathbf{r}_1d\mathbf{r}_2d\mathbf{r}_3d\mathbf{r}_4d\mathbf{r}_5\notag\\
\times&\langle\mathcal{T} E_L^{\dagger}(\omega,\mathbf{r}_a)E_L^{\dagger}(\tau_1,\mathbf{r}_1)E_L(\tau_2,\mathbf{r}_2)E_L(\tau_3,\mathbf{r}_3)E_R(\tau_4,\mathbf{r}_4)E_R^{\dagger}(\tau_5,\mathbf{r}_5)\rangle\notag\\
\times&\langle \mathcal{T}V_L(t,\mathbf{r}_a)V_L(\tau_1,\mathbf{r}_1)V_L^{\dagger}(\tau_2,\mathbf{r}_2)V_L^{\dagger}(\tau_3,\mathbf{r}_3)V_R^{\dagger}(\tau_4,\mathbf{r}_4)V_R(\tau_5,\mathbf{r}_5)\rangle,
\end{align}
\begin{align}\label{eq:S6900}
S_{A7 }(\omega)=\mathcal{I}\frac{2i}{\hbar^6}&\int_{-\infty}^{\infty}dte^{i\omega t}\int_{-\infty}^td\tau_1\int_{-\infty}^{\tau_1}d\tau_2\int_{-\infty}^{t}d\tau_3\int_{-\infty}^{\tau_3}d\tau_4\int_{-\infty}^{\tau_4}d\tau_5\int d\mathbf{r}_1d\mathbf{r}_2d\mathbf{r}_3d\mathbf{r}_4d\mathbf{r}_5\notag\\
\times&\langle \mathcal{T}E_L^{\dagger}(\omega,\mathbf{r}_a)E_L(\tau_1,\mathbf{r}_1)E_L(\tau_2,\mathbf{r}_2)E_R^{\dagger}(\tau_3,\mathbf{r}_3)E_R(\tau_4,\mathbf{r}_4)E_R^{\dagger}(\tau_5,\mathbf{r}_5)\rangle\notag\\
\times&\langle \mathcal{T}V_L(t,\mathbf{r}_a)V_L^{\dagger}(\tau_1,\mathbf{r}_1)V_L^{\dagger}(\tau_2,\mathbf{r}_2)V_R(\tau_3,\mathbf{r}_3)V_R^{\dagger}(\tau_4,\mathbf{r}_4)V_R(\tau_5,\mathbf{r}_5)\rangle,
\end{align}

The above expressions can be recast in Hilbert space

\begin{align}\label{eq:S610}
S_{A1}(\omega)=-\mathcal{I}\frac{2i}{\hbar^6}&\int_{-\infty}^{\infty}dte^{i\omega t}\int_{-\infty}^td\tau_1\int_{-\infty}^{\tau_1}d\tau_2\int_{-\infty}^{\tau_2}d\tau_3\int_{-\infty}^{\tau_3}d\tau_4\int_{-\infty}^{\tau_4}d\tau_5\int d\mathbf{r}_1d\mathbf{r}_2d\mathbf{r}_3d\mathbf{r}_4d\mathbf{r}_5\notag\\
\times&\langle \mathcal{T}E^{\dagger}(\omega,\mathbf{r}_a)E^{\dagger}(\tau_1,\mathbf{r}_1)E(\tau_2,\mathbf{r}_2)E(\tau_3,\mathbf{r}_3)E^{\dagger}(\tau_4,\mathbf{r}_4)E(\tau_5,\mathbf{r}_5)\rangle\notag\\
\times&\langle \mathcal{T}V(t,\mathbf{r}_a)V(\tau_1,\mathbf{r}_1)V^{\dagger}(\tau_2,\mathbf{r}_2)V^{\dagger}(\tau_3,\mathbf{r}_3)V(\tau_4,\mathbf{r}_4)V^{\dagger}(\tau_5,\mathbf{r}_5)\rangle,
\end{align}
\begin{align}
S_{A2}(\omega)=-\mathcal{I}\frac{2i}{\hbar^6}&\int_{-\infty}^{\infty}dte^{i\omega t}\int_{-\infty}^td\tau_1\int_{-\infty}^{\tau_1}d\tau_2\int_{-\infty}^{\tau_2}d\tau_3\int_{-\infty}^{\tau_3}d\tau_4\int_{-\infty}^{\tau_4}d\tau_5\int d\mathbf{r}_1d\mathbf{r}_2d\mathbf{r}_3d\mathbf{r}_4d\mathbf{r}_5\notag\\
\times&\langle \mathcal{T}E^{\dagger}(\omega,\mathbf{r}_a)E(\tau_1,\mathbf{r}_1)E^{\dagger}(\tau_2,\mathbf{r}_2)E^{\dagger}(\tau_3,\mathbf{r}_3)E(\tau_4,\mathbf{r}_4)E(\tau_5,\mathbf{r}_5)\rangle\notag\\
\times&\langle \mathcal{T}V(t,\mathbf{r}_a)V^{\dagger}(\tau_1,\mathbf{r}_1)V(\tau_2,\mathbf{r}_2)V(\tau_3,\mathbf{r}_3)V^{\dagger}(\tau_4,\mathbf{r}_4)V^{\dagger}(\tau_5,\mathbf{r}_5)\rangle,
\end{align}
\begin{align}
S_{A3 }(\omega)=-\mathcal{I}\frac{2i}{\hbar^6}&\int_{-\infty}^{\infty}dte^{i\omega t}\int_{-\infty}^td\tau_1\int_{-\infty}^{\tau_1}d\tau_2\int_{-\infty}^{\tau_2}d\tau_3\int_{-\infty}^{\tau_3}d\tau_4\int_{-\infty}^{\tau_4}d\tau_5\int d\mathbf{r}_1d\mathbf{r}_2d\mathbf{r}_3d\mathbf{r}_4d\mathbf{r}_5\notag\\
&\langle \mathcal{T}E^{\dagger}(\omega,\mathbf{r}_a)E^{\dagger}(\tau_1,\mathbf{r}_1)E(\tau_2,\mathbf{r}_2)E^{\dagger}(\tau_3,\mathbf{r}_3)E(\tau_4,\mathbf{r}_4)E(\tau_5,\mathbf{r}_5)\rangle\notag\\
\times&\langle \mathcal{T}V(t,\mathbf{r}_a)V(\tau_1,\mathbf{r}_1)V^{\dagger}(\tau_2,\mathbf{r}_2)V(\tau_3,\mathbf{r}_3)V^{\dagger}(\tau_4,\mathbf{r}_4)V^{\dagger}(\tau_5,\mathbf{r}_5)\rangle,
\end{align}
\begin{align}
S_{A4 }(\omega)=\mathcal{I}\frac{2i}{\hbar^6}&\int_{-\infty}^{\infty}dte^{i\omega t}\int_{-\infty}^td\tau_1\int_{-\infty}^{\tau_1}d\tau_2\int_{-\infty}^{\tau_2}d\tau_3\int_{-\infty}^{\tau_3}d\tau_4\int_{-\infty}^{t}d\tau_5\int d\mathbf{r}_1d\mathbf{r}_2d\mathbf{r}_3d\mathbf{r}_4d\mathbf{r}_5\notag\\
&\langle \mathcal{T}E^{\dagger}(\tau_5,\mathbf{r}_5)E^{\dagger}(\omega,\mathbf{r}_a)E(\tau_1,\mathbf{r}_1)E(\tau_2,\mathbf{r}_2)E^{\dagger}(\tau_3,\mathbf{r}_3)E(\tau_4,\mathbf{r}_4)\rangle\notag\\
\times&\langle \mathcal{T}V(\tau_5,\mathbf{r}_5)V(t,\mathbf{r}_a)V^{\dagger}(\tau_1,\mathbf{r}_1)V^{\dagger}(\tau_2,\mathbf{r}_2)V(\tau_3,\mathbf{r}_3)V^{\dagger}(\tau_4,\mathbf{r}_4)\rangle,
\end{align}
\begin{align}
S_{A5 }(\omega)=\mathcal{I}\frac{2i}{\hbar^6}&\int_{-\infty}^{\infty}dte^{i\omega t}\int_{-\infty}^td\tau_1\int_{-\infty}^{\tau_1}d\tau_2\int_{-\infty}^{\tau_2}d\tau_3\int_{-\infty}^{\tau_3}d\tau_4\int_{-\infty}^{t}d\tau_5\int d\mathbf{r}_1d\mathbf{r}_2d\mathbf{r}_3d\mathbf{r}_4d\mathbf{r}_5\notag\\
&\langle \mathcal{T}E^{\dagger}(\tau_5,\mathbf{r}_5)E^{\dagger}(\omega,\mathbf{r}_a)E(\tau_1,\mathbf{r}_1)E^{\dagger}(\tau_2,\mathbf{r}_2)E(\tau_3,\mathbf{r}_3)E(\tau_4,\mathbf{r}_4)\rangle\notag\\
\times&\langle \mathcal{T}V(\tau_5,\mathbf{r}_5)V(t,\mathbf{r}_a)V^{\dagger}(\tau_1,\mathbf{r}_1)V(\tau_2,\mathbf{r}_2)V^{\dagger}(\tau_3,\mathbf{r}_3)V^{\dagger}(\tau_4,\mathbf{r}_4)\rangle,
\end{align}
\begin{align}
S_{A6 }(\omega)=-\mathcal{I}\frac{2i}{\hbar^6}&\int_{-\infty}^{\infty}dte^{i\omega t}\int_{-\infty}^td\tau_1\int_{-\infty}^{\tau_1}d\tau_2\int_{-\infty}^{\tau_2}d\tau_3\int_{-\infty}^{t}d\tau_4\int_{-\infty}^{\tau_4}d\tau_5\int d\mathbf{r}_1d\mathbf{r}_2d\mathbf{r}_3d\mathbf{r}_4d\mathbf{r}_5\notag\\
&\langle \mathcal{T}E^{\dagger}(\tau_5,\mathbf{r}_5)E(\tau_4,\mathbf{r}_4)E^{\dagger}(\omega,\mathbf{r}_a)E^{\dagger}(\tau_1,\mathbf{r}_1)E(\tau_2,\mathbf{r}_2)E(\tau_3,\mathbf{r}_3)\rangle\notag\\
\times&\langle \mathcal{T}V(\tau_5,\mathbf{r}_5)V^{\dagger}(\tau_4,\mathbf{r}_4)V(t,\mathbf{r}_a)V(\tau_1,\mathbf{r}_1)V^{\dagger}(\tau_2,\mathbf{r}_2)V^{\dagger}(\tau_3,\mathbf{r}_3)\rangle,
\end{align}
\begin{align}\label{eq:S690}
S_{A7 }(\omega)=\mathcal{I}\frac{2i}{\hbar^6}&\int_{-\infty}^{\infty}dte^{i\omega t}\int_{-\infty}^td\tau_1\int_{-\infty}^{\tau_1}d\tau_2\int_{-\infty}^{t}d\tau_3\int_{-\infty}^{\tau_3}d\tau_4\int_{-\infty}^{\tau_4}d\tau_5\int d\mathbf{r}_1d\mathbf{r}_2d\mathbf{r}_3d\mathbf{r}_4d\mathbf{r}_5\notag\\
&\langle \mathcal{T}E^{\dagger}(\tau_5,\mathbf{r}_5)E(\tau_4,\mathbf{r}_4)E^{\dagger}(\tau_3,\mathbf{r}_3)E^{\dagger}(\omega,\mathbf{r}_a)E(\tau_1,\mathbf{r}_1)E(\tau_2,\mathbf{r}_2)\rangle\notag\\
\times&\langle \mathcal{T}V(\tau_5,\mathbf{r}_5)V^{\dagger}(\tau_4,\mathbf{r}_4)V(\tau_3,\mathbf{r}_3)V(t,\mathbf{r}_a)V^{\dagger}(\tau_1,\mathbf{r}_1)V^{\dagger}(\tau_2,\mathbf{r}_2)\rangle,
\end{align}

Note, that Eqs. (\ref{eq:S610}) - (\ref{eq:S690}) contain non normally ordered correlation functions of the electric field operators and thus contribute to nonclassical features. Since the classical response of a system of noninteracting atoms contains no collective features we shall focus on the nonclassical signal. These equations can be recast in the frequency domain by expanding the field $E(\tau_j,\mathbf{r}_\alpha)=\int_{-\infty}^{\infty}\frac{d\omega_j}{2\pi}E(\omega_j,\mathbf{r}_\alpha)e^{-i\omega_j\tau_j}$ and taking into account that $V(t,\mathbf{r})=\sum_{\alpha}\mathbf{V}^{\alpha}(t)\delta(\mathbf{r}-\mathbf{r}_{\alpha})$, $\alpha=a,b$

\begin{align}\label{eq:S611}
S_{A1}(\omega)=-\mathcal{I}\frac{2i}{\hbar^6}\sum_{\alpha,\beta}\int_{-\infty}^{\infty}\frac{d\omega_1}{2\pi}\frac{d\omega_2}{2\pi}\frac{d\omega_3}{2\pi}\frac{d\omega_4}{2\pi}\frac{d\omega_5}{2\pi}&\{\langle E^{\dagger}(\omega,\mathbf{r}_a)E^{\dagger}(\omega_1,\mathbf{r}_b)E(\omega_2,\mathbf{r}_{\bar{\beta}})E(\omega_5,\mathbf{r}_\alpha)\rangle[E(\omega_3,\mathbf{r}_\beta),E^{\dagger}(\omega_4,\mathbf{r}_\alpha)]\notag\\
+&\langle E^{\dagger}(\omega,\mathbf{r}_a)E^{\dagger}(\omega_1,\mathbf{r}_b)E(\omega_3,\mathbf{r}_\beta)E(\omega_5,\mathbf{r}_\alpha)\rangle[E(\omega_2,\mathbf{r}_{\bar{\beta}}),E^{\dagger}(\omega_4,\mathbf{r}_\alpha)]\}\notag\\
\times&R^{(\alpha\beta)}_{A1}(\omega,\omega_1,\omega_2,\omega_3,\omega_4,\omega_5),
\end{align}
\begin{align}
S_{A2}(\omega)=-\mathcal{I}\frac{2i}{\hbar^6}\sum_{\alpha,\beta}\int_{-\infty}^{\infty}\frac{d\omega_1}{2\pi}\frac{d\omega_2}{2\pi}\frac{d\omega_3}{2\pi}\frac{d\omega_4}{2\pi}\frac{d\omega_5}{2\pi}&\{\langle E^{\dagger}(\omega,\mathbf{r}_a)E^{\dagger}(\omega_2,\mathbf{r}_\beta)E(\omega_4,\mathbf{r}_{\bar{\alpha}})E(\omega_5,\mathbf{r}_\alpha)\rangle[E(\omega_1,\mathbf{r}_a),E^{\dagger}(\omega_3,\mathbf{r}_{\bar{\beta}})]\notag\\
+&\langle E^{\dagger}(\omega,\mathbf{r}_a)E^{\dagger}(\omega_3,\mathbf{r}_{\bar{\beta}})E(\omega_4,\mathbf{r}_{\bar{\alpha}})E(\omega_5,\mathbf{r}_\alpha)\rangle[E(\omega_1,\mathbf{r}_a),E^{\dagger}(\omega_2,\mathbf{r}_\beta)]\}\notag\\
\times&R^{(\alpha\beta)}_{A2}(\omega,\omega_1,\omega_2,\omega_3,\omega_4,\omega_5),
\end{align}
\begin{align}
S_{A3 }(\omega)=-\mathcal{I}\frac{2i}{\hbar^6}\sum_{\alpha,\beta}\int_{-\infty}^{\infty}\frac{d\omega_1}{2\pi}\frac{d\omega_2}{2\pi}\frac{d\omega_3}{2\pi}\frac{d\omega_4}{2\pi}\frac{d\omega_5}{2\pi}&\langle E^{\dagger}(\omega,\mathbf{r}_a)E^{\dagger}(\omega_1,\mathbf{r}_b)E(\omega_4,\mathbf{r}_{\bar{\alpha}})E(\omega_5,\mathbf{r}_\alpha)\rangle[E(\omega_2,\mathbf{r}_{\bar{\beta}}),E^{\dagger}(\omega_3,\mathbf{r}_{\bar{\beta}})]\notag\\
\times&R^{(\alpha\beta)}_{A3}(\omega,\omega_1,\omega_2,\omega_3,\omega_4,\omega_5),
\end{align}
\begin{align}
S_{A4 }(\omega)=\mathcal{I}\frac{2i}{\hbar^6}\sum_{\alpha,\beta}\int_{-\infty}^{\infty}\frac{d\omega_1}{2\pi}\frac{d\omega_2}{2\pi}\frac{d\omega_3}{2\pi}\frac{d\omega_4}{2\pi}\frac{d\omega_5}{2\pi}&\{\langle E^{\dagger}(\omega_5,\mathbf{r}_b)E^{\dagger}(\omega,\mathbf{r}_a)E(\omega_2,\mathbf{r}_\beta)E(\omega_4,\mathbf{r}_\alpha)\rangle[E(\omega_1,\mathbf{r}_{\bar{\beta}}),E^{\dagger}(\omega_3,\mathbf{r}_\alpha)]\notag\\
+&\langle E^{\dagger}(\omega_5,\mathbf{r}_b)E^{\dagger}(\omega,\mathbf{r}_a)E(\omega_1,\mathbf{r}_{\bar{\beta}})E(\omega_4,\mathbf{r}_\alpha)\rangle[E(\omega_2,\mathbf{r}_\beta),E^{\dagger}(\omega_3,\mathbf{r}_\alpha)]\}\notag\\
\times&R^{(\alpha\beta)}_{A4}(\omega,\omega_1,\omega_2,\omega_3,\omega_4,\omega_5),
\end{align}
\begin{align}
S_{A5 }(\omega)=\mathcal{I}\frac{2i}{\hbar^6}\sum_{\alpha,\beta}\int_{-\infty}^{\infty}\frac{d\omega_1}{2\pi}\frac{d\omega_2}{2\pi}\frac{d\omega_3}{2\pi}\frac{d\omega_4}{2\pi}\frac{d\omega_5}{2\pi}&\langle E^{\dagger}(\omega_5,\mathbf{r}_b)E^{\dagger}(\omega,\mathbf{r}_a)E(\omega_3,\mathbf{r}_{\bar{\alpha}})E(\omega_4,\mathbf{r}_\alpha)\rangle[E(\omega_1,\mathbf{r}_{\bar{\beta}}),E^{\dagger}(\omega_2,\mathbf{r}_{\bar{\beta}})]\notag\\
\times&R^{(\alpha\beta)}_{A5}(\omega,\omega_1,\omega_2,\omega_3,\omega_4,\omega_5),
\end{align}
\begin{align}
S_{A6 }(\omega)=-\mathcal{I}\frac{2i}{\hbar^6}\sum_{\alpha,\beta}\int_{-\infty}^{\infty}\frac{d\omega_1}{2\pi}\frac{d\omega_2}{2\pi}\frac{d\omega_3}{2\pi}\frac{d\omega_4}{2\pi}\frac{d\omega_5}{2\pi}&\langle E^{\dagger}(\omega_5,\mathbf{r}_\beta)E^{\dagger}(\omega,\mathbf{r}_a)E(\omega_2,\mathbf{r}_{\bar{\alpha}})E(\omega_3,\mathbf{r}_\alpha)\rangle[E(\omega_4,\mathbf{r}_\beta),E^{\dagger}(\omega_1,\mathbf{r}_b)]\notag\\
\times&R^{(\alpha\beta)}_{A6}(\omega,\omega_1,\omega_2,\omega_3,\omega_4,\omega_5),
\end{align}
\begin{align}\label{eq:S691}
S_{A7 }(\omega)=\mathcal{I}\frac{2i}{\hbar^6}\sum_{\alpha,\beta}\int_{-\infty}^{\infty}\frac{d\omega_1}{2\pi}\frac{d\omega_2}{2\pi}\frac{d\omega_3}{2\pi}\frac{d\omega_4}{2\pi}\frac{d\omega_5}{2\pi}&\langle E^{\dagger}(\omega_5,\mathbf{r}_\beta)E^{\dagger}(\omega,\mathbf{r}_a)E(\omega_1,\mathbf{r}_{\bar{\alpha}})E(\omega_2,\mathbf{r}_\alpha)\rangle[E(\omega_4,\mathbf{r}_\beta),E^{\dagger}(\omega_3,\mathbf{r}_b)]\notag\\
\times&R^{(\alpha\beta)}_{A7}(\omega,\omega_1,\omega_2,\omega_3,\omega_4,\omega_5).
\end{align}
Here

\begin{align}\label{eq:R6A1}
R^{(\alpha\beta)}_{A1}(\omega,\omega_1,\omega_2,\omega_3,\omega_4,\omega_5)=2\pi&|\mu_\alpha|^2\mu_\beta^{*}\mu_{\bar{\beta}}^{*}\mu_A\mu_B\delta(\omega+\omega_1-\omega_2-\omega_3+\omega_4-\omega_5)G_{ab}^{(+)}(\omega+\omega_1)\notag\\
\times &G_a(\omega)G_{\alpha}(\omega_5)G_\beta(\omega+\omega_1-\omega_2)G_g(\omega_5-\omega_4),
\end{align}
\begin{align}\label{eq:R6A2}
R^{(\alpha\beta)}_{A2}(\omega,\omega_1,\omega_2,\omega_3,\omega_4,\omega_5)=2\pi&\mu_{\alpha}^{*}\mu_{\bar{\alpha}}^{*}\mu_{\bar{\beta}}\mu_\beta|\mu_A|^2\delta(\omega-\omega_1+\omega_2+\omega_3-\omega_4-\omega_5)G_{ab}^{(+)}(\omega_4+\omega_5)\notag\\
\times &G_a(\omega)G_\alpha(\omega_5)G_\beta(\omega+\omega_2-\omega_1)G_g(\omega_4+\omega_5-\omega_2-\omega_3),
\end{align}
\begin{align}\label{eq:R6A3}
R^{(\alpha\beta)}_{A3}(\omega,\omega_1,\omega_2,\omega_3,\omega_4,\omega_5)=2\pi &\mu_\alpha^{*}\mu_{\bar{\alpha}}^{*}|\mu_{\bar{\beta}}|^2\mu_A\mu_B\delta(\omega+\omega_1-\omega_2+\omega_3-\omega_4-\omega_5)G_{ab}^{(+)}(\omega_4+\omega_5)G_{ab}^{(+)}(\omega+\omega_1)\notag\\
\times &G_a(\omega)G_{\beta}(\omega+\omega_1-\omega_2)G_\alpha(\omega_5),
\end{align}
\begin{align}\label{eq:R6A4}
R^{(\alpha\beta)}_{A4}(\omega,\omega_1,\omega_2,\omega_3,\omega_4,\omega_5)=2\pi&\mu_\beta^{*}\mu_{\bar{\beta}}^{*}|\mu_\alpha|^2\mu_A\mu_B\delta(\omega-\omega_1-\omega_2+\omega_3-\omega_4+\omega_5)G_{ab}^{(+)}(\omega+\omega_5)\notag\\
\times&G_b^{\dagger}(\omega_5)G_\beta(\omega_2+\omega_4-\omega_3)G_{\alpha}(\omega_4)G_g(\omega_4-\omega_3),
\end{align}
\begin{align}\label{eq:R6A5}
R^{(\alpha\beta)}_{A5}(\omega,\omega_1,\omega_2,\omega_3,\omega_4,\omega_5)=2\pi&\mu_\alpha^{*}\mu_{\bar{\alpha}}^{*}|\mu_{\bar{\beta}}|^2\mu_A\mu_B\delta(\omega-\omega_1+\omega_2-\omega_3-\omega_4+\omega_5)G_{ab}^{(+)}(\omega_3+\omega_4)G_{ab}^{(+)}(\omega+\omega_5)\notag\\
\times &G_b^{\dagger}(\omega_5)G_{\beta}(\omega+\omega_5-\omega_1)G_\alpha(\omega_4),
\end{align}
\begin{align}\label{eq:R6A6}
R^{(\alpha\beta)}_{A6}(\omega,\omega_1,\omega_2,\omega_3,\omega_4,\omega_5)=2\pi&\mu_{\alpha}^{*}\mu_{\bar{\alpha}}^{*}|\mu_\beta|^2\mu_A\mu_B\delta(\omega+\omega_1-\omega_2-\omega_3-\omega_4+\omega_5)G_{ab}^{(+)}(\omega_2+\omega_3)\notag\\
\times &G_a(\omega+\omega_5-\omega_4)G_\beta^{\dagger}(\omega_5)G_\alpha(\omega_3)G_g(\omega_5-\omega_4),
\end{align}
\begin{align}\label{eq:R6A8}
R^{(\alpha\beta)}_{A7}(\omega,\omega_1,\omega_2,\omega_3,\omega_4,\omega_5)=2\pi&\mu_\alpha^{*}\mu_{\bar{\alpha}}^{*}|\mu_\beta|^2\mu_A\mu_B\delta(\omega-\omega_1-\omega_2+\omega_3-\omega_4+\omega_5)G_{ab}^{(+)}(\omega_1+\omega_2)\notag\\
\times &G_b^{\dagger}(-\omega+\omega_1+\omega_2)G_\beta^{\dagger}(\omega_5)G_\alpha(\omega_2)G_g^{\dagger}(\omega_5-\omega_4).
\end{align}
Note that due to permutations $\alpha,\beta=a,b$ a single diagram in Fig. \ref{fig:6g} represents four single quantum pathway. Pathways (\ref{eq:R6A1}) - (\ref{eq:R6A8}) contribute for both SC and QED results that differ by a field correlation function. We also note, that field part consists of a commutator that involves two quantum modes and a four point correlation function of classical field. The latter is simply a product of four classical amplitudes. The former can be calculated using the commutation relation (\ref{eq:com}) - (\ref{eq:com1}). Perfoming the frequency integrations we obtain the signal (\ref{eq:S6f}) with nonlinear susceptibilities given by Eq. (\ref{eq:csc0}) where
\begin{align}\label{eq:csc1}
[\chi_{1LLLL}^{(3)I}+\chi_{4LLLR}^{(3)I}+\chi_{4(L\leftrightarrow R)}^{(3)I}]&(-\omega,-\omega_1,\omega+\omega_1-\omega_2,\omega_2)\notag\\
&=G_{ab}^{(+)}(\omega+\omega_1)[G_s^{\dagger}(\omega_1)-G_s(\omega)]\sum_{\alpha,\beta}\tilde{\mathcal{L}}_{\alpha\beta}(\omega_2)e^{-i\mathbf{k}_0(\mathbf{r}_\alpha-\mathbf{r}_\beta)}G_\alpha(\omega_2)G_\beta(\omega_2),
\end{align}
\begin{align}
\chi_{3LLLL}^{(3)I}(-\omega,-\omega_1,\omega+\omega_1-\omega_2,\omega_2)&+[\chi_{5LLLR}^{(3)I}+\chi_{5(L\leftrightarrow R)}^{(3)I}](-\omega,-\omega_1,\omega+\omega_1-\omega_2,\omega_2)\notag\\
&=G_{ab}^{(+)2}(\omega+\omega_1)[G_s^{\dagger}(\omega_1)-G_s(\omega)]\mathcal{L}_{s}(\omega+\omega_1)G_s(\omega+\omega_1-\omega_2),
\end{align}
\begin{align}
\chi_{2LLLL}^{(3)I}(-\omega,-\omega_1,\omega+\omega_1-\omega_2,\omega_2)=-G_{ab}^{(+)}(\omega+\omega_1)G_s(\omega+\omega_1-\omega_2)\sum_{\alpha,\beta}\mathcal{L}_{\alpha\beta}(\omega)e^{-i\mathbf{k}_0(\mathbf{r}_\alpha-\mathbf{r}_\beta)}G_\alpha(\omega)G_\beta(\omega),
\end{align}
\begin{align}\label{eq:csc7}
[\chi_{7LLLR}^{(3)I}+\chi_{7(L\leftrightarrow R)}^{(3)I}](-\omega,-\omega_1,\omega+\omega_1-\omega_2,\omega_2)=G_{ab}^{(+)}(\omega+\omega_1)G_s(\omega+\omega_1-\omega_2)\sum_{\alpha,\beta}\mathcal{L}_{\alpha\beta}(\omega_1)e^{-i\mathbf{k}_0(\mathbf{r}_\alpha-\mathbf{r}_\beta)}G_\alpha^{\dagger}(\omega_1)G_\beta^{\dagger}(\omega_1),
\end{align}
where the couplings $\mathcal{L}_{\alpha\beta}(\omega)$ and $\mathcal{L}_{s}(\omega+\omega_1)$ are given by Eqs. (\ref{eq:Lc}) - (\ref{eq:Ls}), respectively, $\tilde{\mathcal{L}}_{\alpha\beta}(\omega)=\mathcal{L}_{\alpha\beta}(\omega)|\mu_\beta|^2/|\mu_\alpha|^2$, and $G_s(\omega)=G_a(\omega)+G_b(\omega)$. Similarly we evaluate the QED contribution to the susceptibility we get
\begin{align}\label{eq:cq1}
[\chi_{1LLLLLL}^{(5)II}+\chi_{4LLLLLR}^{(5)II}+\chi_{4(L\leftrightarrow R)}^{(5)II}]&(-\omega,-\omega_1,\omega',\omega+\omega_1-\omega_2,-\omega',\omega_2)=G_{ab}^{(+)}(\omega+\omega_1)[G_s^{\dagger}(\omega_1)-G_s(\omega)]\notag\\
\times\sum_{\alpha,\beta}&e^{-i\mathbf{k}_0(\mathbf{r}_\alpha-\mathbf{r}_\beta)}\mu_\beta^{(l)*}\mu_{\alpha}^{(m)}\mathcal{D}_{\alpha\beta}^{(l,m)}(\omega')G_g(\omega_2-\omega')G_\alpha(\omega_2)G_\beta(\omega+\omega_1-\omega'),
\end{align}
\begin{align}
\chi_{2LLLLLL}^{(5)II}(-\omega,\omega',-\omega_1,-\omega',\omega+\omega_1-&\omega_2,\omega_2)=-G_{ab}^{(+)}(\omega+\omega_1)G_s(\omega+\omega_1-\omega_2)\notag\\
\times\sum_{\alpha,\beta}&e^{-i\mathbf{k}_0(\mathbf{r}_\alpha-\mathbf{r}_\beta)}\mu_\alpha^{(l)*}\mu_{\beta}^{(m)}\mathcal{D}_{\alpha\beta}^{(l,m)}(\omega')G_g(\omega-\omega')\mathcal{L}_{\alpha\beta}(\omega)G_\alpha(\omega)G_\beta(\omega+\omega_1-\omega'),
\end{align}
\begin{align}\label{eq:cq6}
[\chi_{6LLLLRR}^{(5)II}+\chi_{6(L\leftrightarrow R)}^{(5)II}](-\omega,\omega',-\omega_1,&-\omega',\omega+\omega_1-\omega_2,\omega_2)=-G_{ab}^{(+)}(\omega+\omega_1)G_s(\omega_2)\notag\\
\times\sum_{\alpha,\beta}&e^{-i\mathbf{k}_0(\mathbf{r}_\alpha-\mathbf{r}_\beta)}\mu_\alpha^{(l)*}\mu_{\beta}^{(m)}\mathcal{D}_{\alpha\beta}^{(l,m)}(\omega')G_g(\omega_1-\omega')G_\alpha^{\dagger}(\omega_1)G_\beta(\omega+\omega_1-\omega').
\end{align}
In the absence of the bath, assuming the ground state frequency and linewidth to be zero $\omega_g=0$, $\gamma_g=0$, we have  $G_g(\omega)\simeq\delta(\omega)$.

%Eqs. (\ref{eq:cq1}) - (\ref{eq:cq6}) then yield
%\begin{align}\label{eq:cq14}
%\sum_{j=1,4}\chi_{IIj}^{(3)}(-\omega,-\omega_1,\omega+\omega_1-\omega_2,\omega_2)=G_{ab}^{(+)}(\omega+\omega_1)[G_s^{\dagger}(\omega_1)-G_s(\omega)]\sum_{\alpha,\beta}\tilde{\mathcal{L}}_{\alpha\beta}(\omega_2)e^{-i\mathbf{k}_0(\mathbf{r}_\alpha-\mathbf{r}_\beta)}G_\alpha(\omega_2)G_\beta(\omega+\omega_1-\omega_2),
%\end{align}
%\begin{align}
%\chi_{II2}^{(5)}(-\omega,-\omega_1,\omega+\omega_1-\omega_2,\omega_2)=-G_{ab}^{(+)}(\omega+\omega_1)G_s(\omega+\omega_1-\omega_2)\sum_{\alpha,\beta}\mathcal{L}_{\alpha\beta}(\omega)e^{-i\mathbf{k}_0(\mathbf{r}_\alpha-\mathbf{r}_\beta)}G_\alpha(\omega)G_\beta(\omega_1),
%\end{align}
%\begin{align}\label{eq:cq61}
%\chi_{II6}^{(3)}(-\omega,-\omega_1,\omega+\omega_1-\omega_2,\omega_2)=-G_{ab}^{(+)}(\omega+\omega_1)G_s(\omega_2)\sum_{\alpha,\beta}\mathcal{L}_{\alpha\beta}(\omega_1)e^{-i\mathbf{k}_0(\mathbf{r}_\alpha-\mathbf{r}_\beta)}G_\alpha^{\dagger}(\omega_1)G_\beta(\omega).
%\end{align}

\section{The transmission of shaped pulses}\label{app:par}

Substituting susceptibilities (\ref{eq:csc1}) - (\ref{eq:cq6}) into the signal (\ref{eq:S6f}) and utilizing the pulse shaping in the field correlation function (\ref{eq:Ecor}) we obtain for the semiclassical contribution Eq. (\ref{eq:S6sc}) where
\begin{align}
A_1(\omega,\omega_p)=&[G_s^{\dagger}(\omega_p)-G_s(\omega)]\int_{-\infty}^{\infty}d\omega_2e^{i[\phi(\omega+\omega_p-\omega_2)+\phi(\omega_2)-\phi(\omega)-\xi]}\sum_{\alpha,\beta}\tilde{\mathcal{L}}_{\alpha\beta}(\omega_2)e^{-i\mathbf{k}_0(\mathbf{r}_\alpha-\mathbf{r}_\beta)}G_\alpha(\omega_2)G_\beta(\omega_2)\notag\\
-&2\pi\sum_{\alpha}e^{i[\phi(\omega_\alpha-i\gamma_\alpha)+\phi(\omega+\omega_p-\omega_\alpha+i\gamma_\alpha)-\phi(\omega)-\xi]}\sum_{\beta,\delta}e^{-i\mathbf{k}_0(\mathbf{r}_\beta-\mathbf{r}_\delta)}[\mathcal{L}_{\beta\delta}(\omega)G_\beta(\omega)G_\delta(\omega)-\mathcal{L}_{\beta\delta}(\omega_p)G_\beta^{\dagger}(\omega_p)G_\delta^{\dagger}(\omega_p)],
\end{align}
\begin{align}
A_2(\omega,\omega_p)=2\pi[G_s^{\dagger}(\omega_p)-G_s(\omega)]\sum_{\alpha}e^{i[\phi(\omega_\alpha-i\gamma_\alpha)+\phi(\omega+\omega_p-\omega_\alpha+i\gamma_\alpha)-\phi(\omega)-\xi]}\mathcal{L}_s(\omega+\omega_p),
\end{align}
\begin{align}
A_3^{(\alpha)}(\omega,\omega_p)=2\pi e^{i[\phi(\omega-\omega_p+\omega_\alpha+i\gamma_\alpha)+\xi-\phi(\omega_\alpha+i\gamma_\alpha)-\phi(\omega)]}G_{\bar{\alpha}}^2(\omega)\mathcal{L}_s(\omega+\omega_\alpha+i\gamma_\alpha),
\end{align}
\begin{align}
A_4^{(\alpha\beta\gamma)}(\omega,\omega_p)=2\pi e^{i[\phi(\omega-\omega_p+\omega_\alpha+i\gamma_\alpha)+\xi-\phi(\omega_\alpha+i\gamma_\alpha)-\phi(\omega)]}G_{\bar{\alpha}}(\omega)\mathcal{M}_{\beta\delta}(\omega-\omega_p+\omega_\alpha+i\gamma_\alpha)e^{-i\mathbf{k}_0(\mathbf{r}_\beta-\mathbf{r}_\delta)},
\end{align}
\begin{align}
A_5^{(\alpha\beta\gamma)}(\omega,\omega_p)=2\pi e^{i[-\phi(\omega_p-\omega+\omega_\alpha-i\gamma_\alpha)+\xi+\phi(\omega_\alpha-i\gamma_\alpha)-\phi(\omega)]}G_{\bar{\alpha}}(\omega_p)\mathcal{M}_{\beta\delta}(\omega_p-\omega+\omega_\alpha-i\gamma_\alpha)e^{-i\mathbf{k}_0(\mathbf{r}_\beta-\mathbf{r}_\delta)},
\end{align}
\begin{align}
A_6(\omega,\omega_p)=&[G_s^{\dagger}(\omega_p)-G_s(\omega)]\sum_{\alpha,\beta}\tilde{\mathcal{L}}_{\alpha\beta}(\omega)e^{-i\mathbf{k}_0(\mathbf{r}_\alpha-\mathbf{r}_\beta)}G_\alpha(\omega)G_\beta(\omega)\notag\\
+&[G_s(\omega)+G_s(\omega_p)][\sum_{\alpha,\beta}\mathcal{L}_{\alpha\beta}(\omega_p)e^{-i\mathbf{k}_0(\mathbf{r}_\alpha-\mathbf{r}_\beta)}G_\alpha^{\dagger}(\omega_p)G_\beta^{\dagger}(\omega_p)-1]
\end{align}
\begin{align}
A_7(\omega,\omega_p)=&e^{i[2\xi-\phi(\omega)-\phi(2\omega_p-\omega)]}[G_s^{\dagger}(2\omega_p-\omega)-G_s(\omega)]\sum_{\alpha,\beta}\tilde{\mathcal{L}}_{\alpha\beta}(\omega)e^{-i\mathbf{k}_0(\mathbf{r}_\alpha-\mathbf{r}_\beta)}G_\alpha(\omega)G_\beta(\omega)\notag\\
+&e^{i[2\xi-\phi(\omega)-\phi(2\omega_p-\omega)]}G_s(2\omega_p-\omega)[\sum_{\alpha,\beta}\mathcal{L}_{\alpha\beta}(2\omega_p-\omega)e^{-i\mathbf{k}_0(\mathbf{r}_\alpha-\mathbf{r}_\beta)}G_\alpha^{\dagger}(2\omega_p-\omega)G_\beta^{\dagger}(2\omega_p-\omega)-1],
\end{align}
\begin{align}
A_8(\omega,\omega_p)=[G_s^{\dagger}(\omega_p)-G_s(\omega)][G_s(\omega)+G_s(\omega_p)]L_s(\omega+\omega_p),
\end{align}
\begin{align}
A_9(\omega,\omega_p)=[G_s^{\dagger}(2\omega_p-\omega)-G_s(\omega)]G_s(2\omega_p-\omega)\mathcal{L}_s(2\omega_p)e^{i[2\xi-\phi(\omega)-\phi(2\omega_p-\omega)]}.
\end{align}

Similarly for the QED contribution we obtain Eq. (\ref{eq:S6q}) where
\begin{align}
B_1(\omega,\omega_p)=-2\pi\sum_{\alpha,\beta,\delta}e^{i[\phi(\omega+\omega_p-\omega_\alpha+i\gamma_\alpha)+\phi(\omega_\alpha-i\gamma_\alpha)-\phi(\omega)-\xi]}\mathcal{L}_{\beta\delta}(\omega)e^{-i\mathbf{k}_0(\mathbf{r}_\beta-\mathbf{r}_\delta)}[G_\beta(\omega)G_{\bar{\delta}}(\omega_p)+G_\beta^{\dagger}(\omega_p)G_{\bar{\delta}}(\omega)]
\end{align}
\begin{align}
B_2(\omega,\omega_p)=2\pi[G_s^{\dagger}(\omega_p)-G_s(\omega)]\sum_{\alpha}\tilde{\mathcal{L}}_{\alpha\alpha}(\omega_\alpha-i\gamma_\alpha)e^{i[\phi(\omega+\omega_p-\omega_\alpha+i\gamma_\alpha)+\phi(\omega_\alpha-i\gamma_\alpha)-\phi(\omega)-\xi]}
\end{align}
\begin{align}
B_3(\omega,\omega_p)=2\pi[G_s^{\dagger}(\omega_p)-G_s(\omega)]\tilde{\mathcal{M}}_{ba}(\omega_a-i\gamma_a)e^{-i\mathbf{k}_0(\mathbf{r}_b-\mathbf{r}_a)}e^{i[\phi(\omega+\omega_p-\omega_a+i\gamma_a)+\phi(\omega_a-i\gamma_a)-\phi(\omega)-\xi]}
\end{align}
\begin{align}
B_4(\omega,\omega_p)=2\pi[G_s^{\dagger}(\omega_p)-G_s(\omega)]\tilde{\mathcal{M}}_{ab}(\omega_b-i\gamma_b)e^{-i\mathbf{k}_0(\mathbf{r}_a-\mathbf{r}_b)}e^{i[\phi(\omega+\omega_p-\omega_b+i\gamma_b)+\phi(\omega_b-i\gamma_b)-\phi(\omega)-\xi]}
\end{align}
\begin{align}
B_5^{(\alpha\beta)}(\omega,\omega_p)&=2\pi G_{\bar{\alpha}}(\omega)e^{i[\phi(\omega-\omega_p+\omega_\alpha+i\gamma_\alpha)+\xi-\phi(\omega)-\phi(\omega_\alpha+i\gamma_\alpha)]}\notag\\
&\times\{[\tilde{\mathcal{L}}_{\bar{\beta}\bar{\beta}}(\omega_p)+\tilde{\mathcal{M}}_{\beta\beta}(\omega-\omega_p+\omega_\alpha+i\gamma_\alpha)]G_{\bar{\beta}}(\omega_p)+[\tilde{\mathcal{L}}_{\bar{\beta}\beta}(\omega_p)+\tilde{\mathcal{M}}_{\bar{\beta}\beta}(\omega-\omega_p+\omega_\alpha+i\gamma_\alpha)e^{-i\mathbf{k}_0(\mathbf{r}_{\bar{\beta}}-\mathbf{r}_\beta)}]G_\beta(\omega_p)\}\notag\\
&-2\pi G_{\bar{\alpha}}(\omega)e^{i[\phi(\omega-\omega_p+\omega_\alpha-i\gamma_\alpha)+\xi-\phi(\omega)-\phi(\omega_\alpha-i\gamma_\alpha)]}[\mathcal{L}_{\bar{\alpha}\bar{\alpha}}(\omega)G_{\bar{\alpha}}(\omega)+\mathcal{L}_{\alpha\bar{\alpha}}(\omega)G_\alpha(\omega)]
\end{align}
\begin{align}
B_6^{(\alpha\beta)}(\omega,\omega_p)&=-2\pi G_{\bar{\alpha}}(\omega_p)e^{i[-\phi(\omega_p-\omega+\omega_\alpha-i\gamma_\alpha)+\phi(\omega_\alpha-i\gamma_\alpha)+\xi-\phi(\omega)]}[\mathcal{L}_{\bar{\beta}\bar{\beta}}(\omega)G_{\bar{\beta}}(\omega)+\mathcal{L}_{\beta\bar{\beta}}(\omega)e^{-i\mathbf{k}_0(\mathbf{r}_\beta-\mathbf{r}_{\bar{\beta}})}G_{\beta}(\omega)]
\end{align}
\begin{align}
B_7^{(\alpha\beta)}(\omega,\omega_p)=&-2\pi G_{\bar{\alpha}}(\omega_p)e^{i[-\phi(\omega_p-\omega+\omega_\alpha-i\gamma_\alpha)+\phi(\omega_\alpha-i\gamma_\alpha)+\xi-\phi(\omega)]}\notag\\
\times &[\mathcal{M}_{\beta\beta}(\omega_p-\omega+\omega_\alpha-i\gamma_\alpha)G_{\bar{\beta}}(\omega)+\mathcal{M}_{\beta\bar{\beta}}(\omega_p-\omega+\omega_\alpha-i\gamma_\alpha)e^{-i\mathbf{k}_0(\mathbf{r}_\beta-\mathbf{r}_{\bar{\beta}})}G_{\beta}(\omega)]
\end{align}
\begin{align}
B_8(\omega,\omega_p)=&[G_s^{\dagger}(\omega_p)-G_s(\omega)]\sum_{\alpha,\beta}[\tilde{\mathcal{L}}_{\alpha\beta}(\omega)G_\alpha(\omega)G_{\bar{\beta}}(\omega_p)+\tilde{\mathcal{L}}_{\alpha\beta}(\omega_p)G_{\bar{\alpha}\beta}(\omega)G_\beta(\omega_p)]e^{-i\mathbf{k}_0(\mathbf{r}_\alpha-\mathbf{r}_\beta)}\notag\\
-&[G_s(\omega_p)+G_s(\omega)]\sum_{\alpha,\beta}[\mathcal{L}_{\alpha\beta}(\omega)G_\alpha(\omega)G_{\bar{\beta}}(\omega_p)+\mathcal{L}_{\alpha\beta}(\omega_p)G_\alpha^{\dagger}(\omega_p)G_{\bar{\beta}}(\omega)]e^{-i\mathbf{k}_0(\mathbf{r}_\alpha-\mathbf{r}_\beta)}
\end{align}
\begin{align}
B_9(\omega,\omega_p)=&e^{i[2\xi-\phi(\omega)-\phi(2\omega_p-\omega)]}\{[G_s^{\dagger}(2\omega_p-\omega)-G_s(\omega)]\sum_{\alpha,\beta}\tilde{\mathcal{L}}_{\alpha\beta}(\omega)e^{-i\mathbf{k}_0(\mathbf{r}_\alpha-\mathbf{r}_\beta)}G_{\bar{\alpha}}(2\omega_p-\omega)G_{\beta}(\omega)\notag\\
-&G_s(2\omega_p-\omega)\sum_{\alpha,\beta}[\mathcal{L}_{\alpha\beta}(\omega)G_\alpha(\omega)G_{\bar{\beta}}(2\omega_p-\omega)+\mathcal{L}_{\alpha\beta}(2\omega_p-\omega)G_{\alpha}^{\dagger}(2\omega_p-\omega)G_{\bar{\beta}}(\omega)]e^{-i\mathbf{k}_0(\mathbf{r}_\alpha-\mathbf{r}_\beta)}\}.
\end{align}

In the above expressions the couplings are given by Eqs. (\ref{eq:Lab}) - (\ref{eq:Mab}) and  $\tilde{\mathcal{L}}_{\alpha\beta}(\omega)=\mathcal{L}_{\alpha\beta}(\omega)|\mu_\beta|^2/|\mu_\alpha|^2$ and $\tilde{\mathcal{M}}_{\alpha\beta}(\omega)=\mathcal{M}_{\alpha\beta}(\omega)|\mu_\beta|^2/|\mu_\alpha|^2$.

\end{widetext}

%\nocite{marx,roslyak,mukamelbook,gelmukhanov,biggs,dorfman,scully}
%\bibliography{SLEbib}
%\bibliographystyle{plain}
%

\end{document}